\documentclass[reprint,prb,aps,twocolumn,superscriptaddress, 10pt]{revtex4-1}
\usepackage{amsmath}
\usepackage[english]{babel}
\usepackage{graphicx}
\usepackage{color}
\usepackage[dvipsnames]{xcolor}

\usepackage[colorlinks,
            citecolor = blue, 
						linkcolor = blue,
						urlcolor  = blue,
						anchorcolor = blue,
						breaklinks = true
						]{hyperref}

\usepackage{soul}
\usepackage{ amssymb }

\newif\ifdraft
\drafttrue

\def \ETH{Institute for Quantum Electronics, ETH Z\"urich, CH-8093 Z\"urich, Switzerland}
\def \HARVARD{Department of Physics, Harvard University, Cambridge, Massachusetts 02138, USA}
\def \NIMSRCFM{Research Center for Functional Materials, National Institute for Materials Science, Tsukuba, Ibaraki 305-0044, Japan}
\def \NIMSICMN{International Center for Materials Nanoarchitectonics, National Institute for Materials Science, Tsukuba, Ibaraki 305-0044, Japan}

\def \MCQST{M\"unchen Center for Quantum Science and Technology, Schellingstrasse 4, 80799 M\"unich, Germany}
\def \TUM{Department of Physics and Institute for Advanced Study, Technical University of Munich, 85748 Garching, Germany}

\begin{document}


\title{Observation of Wigner crystal of electrons in a monolayer semiconductor}

\author{T. Smole\'nski}
\affiliation{\ETH}

\author{P. E. Dolgirev}
\affiliation{\HARVARD}

\author{C. Kuhlenkamp}
\affiliation{\ETH}
\affiliation{\TUM}
\affiliation{\MCQST}

\author{A. Popert}
\affiliation{\ETH}

\author{Y. Shimazaki}
\affiliation{\ETH}

\author{P.~Back}
\affiliation{\ETH}

\author{M. Kroner}
\affiliation{\ETH}

\author{K.~Watanabe}
\affiliation{\NIMSRCFM}

\author{T.~Taniguchi}
\affiliation{\NIMSICMN}

\author{I. Esterlis}
\affiliation{\HARVARD}

\author{E. Demler}
\affiliation{\HARVARD}

\author{A. Imamo\u{g}lu}
\affiliation{\ETH}

\maketitle


{\bf When the Coulomb repulsion between electrons dominates over their kinetic energy, electrons in two dimensional systems were predicted to spontaneously break continuous translation symmetry and form a quantum crystal~\cite{Wigner1934}. Efforts to observe\cite{Lozovik1975, Grimes_PRL_1979, Andrei1988, Goldman1990, Williams_PRL_1991, Buhmann1991, Goldys1992, Ye2002, Chen2003, Chen_NatPhys_2006, Tiemann2014, Deng2016} this elusive state of matter, termed a Wigner crystal (WC), in two dimensional extended systems have primarily focused on electrons confined to a single Landau level at high magnetic fields, but have not provided a conclusive experimental signature of the emerging charge order. Here, we use optical spectroscopy to demonstrate that electrons in a pristine monolayer semiconductor with density $ \lesssim 3 \cdot 10^{11}$~cm$^{-2}$ form a WC. The interactions between resonantly injected excitons and electrons arranged in a periodic lattice modify the exciton band structure so that it exhibits a new umklapp resonance, heralding the presence of charge order~\cite{Shimazaki_arXiv_2020}. Remarkably, the combination of a relatively high electron mass and reduced dielectric screening allows us to observe an electronic WC state even in the absence of magnetic field. The tentative phase diagram obtained from our Hartree-Fock calculations provides an explanation of the striking experimental signatures obtained up to $B = 16$~T. Our findings demonstrate that charge-tunable transition metal dichalcogenide (TMD) monolayers~\cite{Mak_PRL_2010,Xu2014} enable the investigation of previously uncharted territory for many-body physics where interaction energy dominates over kinetic energy, even in the absence of a moire potential or external fields.
}


The electronic properties of most metals and semiconductors at low temperatures can be described using the Fermi liquid theory. This is a consequence of the fact that in most material systems typical kinetic energy of electrons exceeds the Coulomb interaction energy. Investigation of strong electronic correlations that emerge in the complementary regime where the ratio~$r_s$ of the Coulomb interactions to the kinetic energy well exceeds unity has been a holy grail of condensed-matter physics. A landmark state of matter that was predicted to appear in this latter limit is a Wigner crystal (WC)~\cite{Wigner1934} where electrons spontaneously break translational symmetry and form a periodic lattice. Quantum Monte Carlo calculations~\cite{Drummond2009} indicate that $r_s \gtrsim 30$ is necessary for the WC to be the ground state of a two dimensional electron system (2DES) when $B=0$. Since $r_s = m_e^* e^2/( 4\pi\epsilon_0\epsilon \hbar^2 \sqrt{\pi n_e})$ with $\epsilon$, $n_e$ and $m_e^*$ denoting the dielectric constant, electron density and effective electron mass, simple considerations show that high quality materials with weak disorder scattering, large $m_e^*$, low $n_e$ and small $\epsilon$ should be used to achieve the requisite $r_s$ values. 

The difficulty in simultaneously satisfying the above mentioned stringent conditions has, with a few remarkable exceptions~\cite{Yoon1999, Shapir2019}, hindered the search for an electronic WC state in the absence of magnetic fields. Instead, majority of the experimental efforts\cite{Lozovik1975, Grimes_PRL_1979, Andrei1988, Goldman1990, Williams_PRL_1991, Buhmann1991, Goldys1992, Ye2002, Chen2003, Chen_NatPhys_2006, Tiemann2014, Deng2016} to date have focused on 2DES in conventional semiconductors under large out-of-plane field $B>0$~: in the limit where $n_e$ is much lower than the Landau level (LL) degeneracy, the kinetic energy of electrons is quenched and the Coulomb interaction provides the only relevant energy scale. While this system is prima facie ideal for electron crystallization, strong dielectric screening, finite disorder, and competition with fractional quantum Hall states have impeded an unequivocal observation of the charge order in the WC state.

\begin{figure*}[t]
	\includegraphics[width=0.99\textwidth]{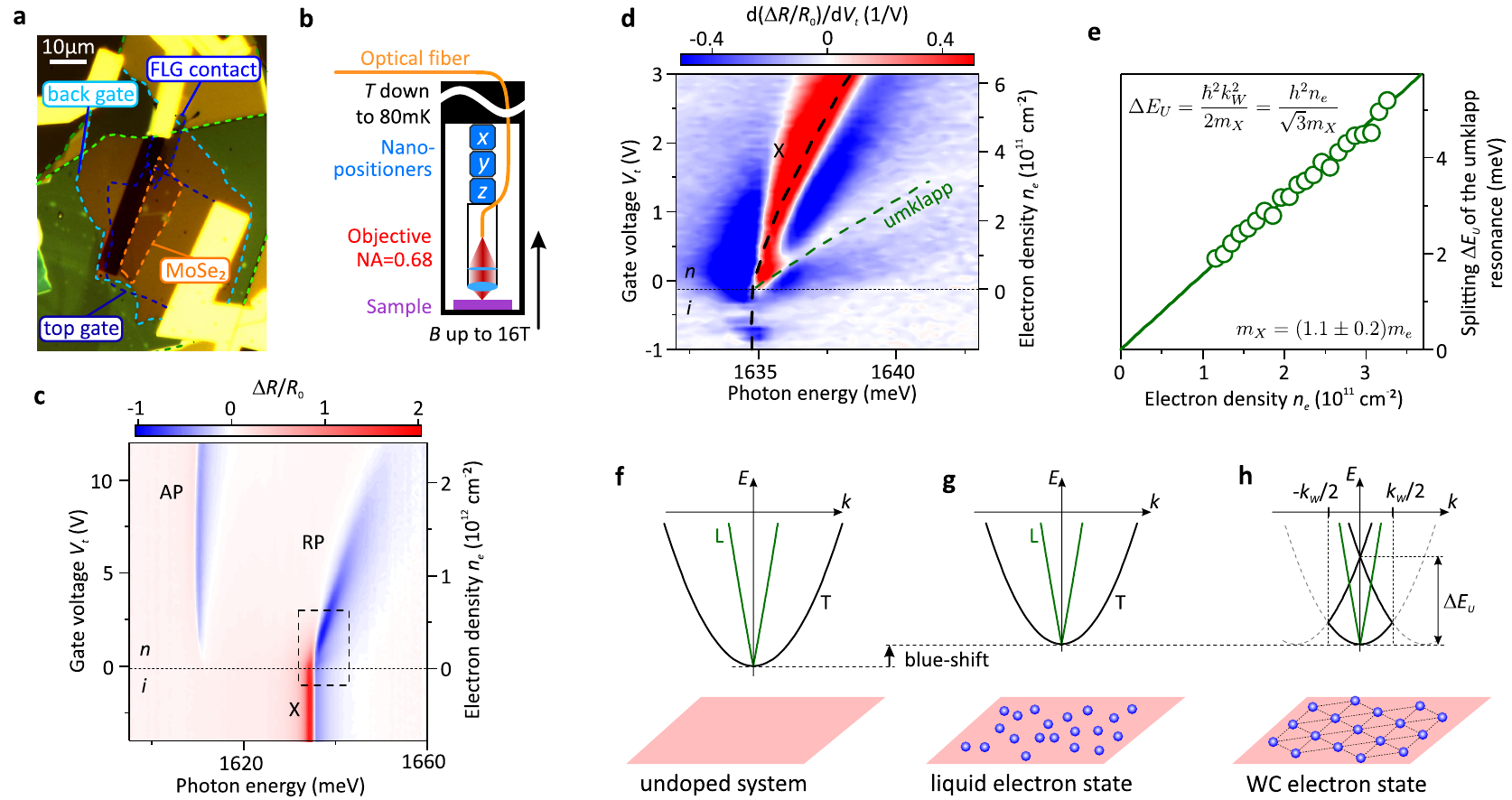}
	\caption{{\bf Optical signatures of a Wigner crystal at zero magnetic field.} ({\bf a})~Optical micrograph of the studied device consisting of a charge-tunable, dual-graphene-gated, and fully-hBN-encapsulated MoSe$_2$ monolayer (dashed lines indicate boundaries of the flakes). ({\bf b})~Schematic of the fiber-based setup for magneto-optical spectroscopy of the device at milikelvin temperatures. ({\bf c})~Gate-voltage-dependence of the reflectance contrast spectrum featuring charge-neutral ($i$) and electron-doped regions ($n$). ({\bf d})~Close-up into the exciton spectral range in the low-density regime (marked by dashed rectangle in panel~{\bf c})~showing derivative of reflectance contrast with respect to the gate voltage. The weak, higher-energy resonance is due to umklapp scattering of the excitons off the electron WC. Black dashed line marks the fitted position $E_X$ of the exciton resonance, while green line corresponds to the expected position of the umklapp peak $E_X+\Delta E_U$. ({\bf e})~Energy splitting $\Delta E_U$ between the exciton and umklapp peaks determined as a function of the electron density $n_e$. Solid line marks the linear fit corresponding to the exciton mass of $m_X=(1.1\pm0.2)m_e$. ({\bf f-h})~Schematics illustrating the exciton dispersion in a monolayer semiconductor hosting an electron system in various structural phases. The exciton bands are split by the electron-hole exchange interaction into parabolic- and linear-in-momentum branches that correspond to the exciton dipole oriented along transverse~(T) or longitudinal~(L) directions with respect to the momentum vector. For the electrons in a liquid state~({\bf g}), the bands are simply blueshifted with respect to the undoped case~({\bf f}). In a WC phase~({\bf h}), the exciton umklapp scattering off the periodic electron lattice leads to band-folding. This gives rise to emergence of a new, zero-momentum umklapp resonance with an energy $\Delta E_U \simeq \hbar^2 k_W^2/2m_X$ determined by reciprocal lattice vector $k_W$. \label{fig:Fig1}}
\end{figure*}

Transition metal dichalcogenide (TMD) monolayers have the promise to overcome this conundrum~\cite{zarenia2017wigner}: the combination of weak dielectric screening together with $m_e^*$ that is an order of magnitude larger than that of GaAs shows that $r_s > 30$ can be obtained for an electron density of $n_e \sim 1 \cdot 10^{11}$~cm$^{-2}$ even in the $B = 0$ limit. The stark contrast to conventional materials becomes even more prominent in the limit $B \neq 0$, where the relevant parameter quantifying the strength of interactions is given by the ratio of the characteristic Coulomb interaction energy ($E_c$) to the cyclotron energy ($\hbar \omega_c$): $E_c/\hbar \omega_c =  m_e^* e / (4\pi\epsilon_0\epsilon l_0 \hbar B) = r_s \sqrt{\nu/2}$, with $l_0 = \sqrt{\hbar/(e B)}$ and $\nu$ denoting the magnetic length and LL filling factor, respectively. Since $E_c/\hbar \omega_c \ge 25$ for $B\sim4$~T in TMD monolayers, interactions lead to strong LL mixing and it is no longer possible to treat electron motion as being restricted to a single LL. 

Here, we report direct evidence for electronic WC in two different monolayer TMD devices. In contrast to the recently observed Wigner-Mott states in moire superlattices~\cite{Wang_Nature_2020,Mak_Nature_2020, Shimazaki2020, Wang_NatMater_2020, Mak_2020_arxiv}, the WC states in our devices spontaneously break continuous translational symmetry. Both samples we investigated consist of a charge-tunable MoSe$_2$ monolayer that was encapsulated between two layers of hexagonal boron-nitride (hBN) and covered with few-layer-graphene flakes serving as top and back transparent gate electrodes [see section S1 of the Supplementary Information (SI) for details regarding sample fabrication and experimental setup]. In the main text we focus on the device depicted in Fig.~\ref{fig:Fig1}{\bf a}, for which the doping was controlled by applying a top gate voltage~$V_{t}$. The resulting electron density $n_e(V_{t})$ was precisely calibrated based on the LL fan chart of the Shubnikov-de Haas oscillations of the exciton linewidth~\cite{Smolenski2019} revealed by our high-magnetic-field studies (section S4 of the SI). This device was mounted in a dilution refrigerator with a monomode-fiber-based optical access (Fig.~\ref{fig:Fig1}{\bf b}) allowing to perform circular-polarization-resolved, magneto-optical experiments with sub-micron spatial resolution at a base temperature of 80~mK (see section S5 of the SI for the reference data sets taken at $T=4$~K). 

We first concentrate on the fingerprints of the WC phase for $B=0$. Fig.~\ref{fig:Fig1}{\bf c} shows representative gate-voltage-evolution of the resonant reflectance contrast spectrum $(R-R_0)/R_0=\Delta R/R_0$ of our device, where $R$ is the reflectance signal measured in the MoSe$_2$-monolayer region, while $R_0$ represents a background reflectance taken off the MoSe$_2$ flake (see section S2 of the SI). At negative $V_t$, when the MoSe$_2$ is charge-neutral, the spectrum displays a single resonance related to the bare exciton (X). As evidenced by several previous reports~\cite{Sidler2017}, upon electron doping (at $V_t>0$), the excitons become dressed into exciton-polarons, due to dynamical screening by the electron system. This leads to emergence of a second, red-shifted resonance, termed the attractive polaron (AP), which gets brighter as $n_e$ is increased. In parallel, the exciton peak smoothly transforms into a repulsive polaron (RP) that blueshifts, becomes broader, and loses its oscillator strength.

In search for signatures of the periodic charge order, we examine the limit of low $n_e$ in the spectral vicinity of the RP transition, which, for simplicity, we will further refer to as the exciton resonance. Remarkably, in this regime we observe a second, higher-energy resonance. Due to its small oscillator strength, about two orders of magnitude lower than that of the exciton, this resonance is barely visible in the raw reflectance spectra, but becomes prominent after differentiating $\Delta R/R_0$ with respect to the gate voltage, or equivalently $n_e$ (Fig.~\ref{fig:Fig1}{\bf d}). This resonance blueshifts faster than the exciton with increasing~$n_e$ and becomes indiscernible at $n_e\gtrsim3\cdot10^{11}\ \mathrm{cm}^{-2}$. At the same time, its energy splitting $\Delta E_U$ from the main exciton transition, determined by fitting both resonances with dispersive Lorentzian spectral profiles (following a procedure described in section S3 of the SI), scales linearly with $n_e$ and extrapolates to zero at $n_e=0$ (Fig.~\ref{fig:Fig1}{\bf e}).

\begin{figure}[t!]
    \includegraphics{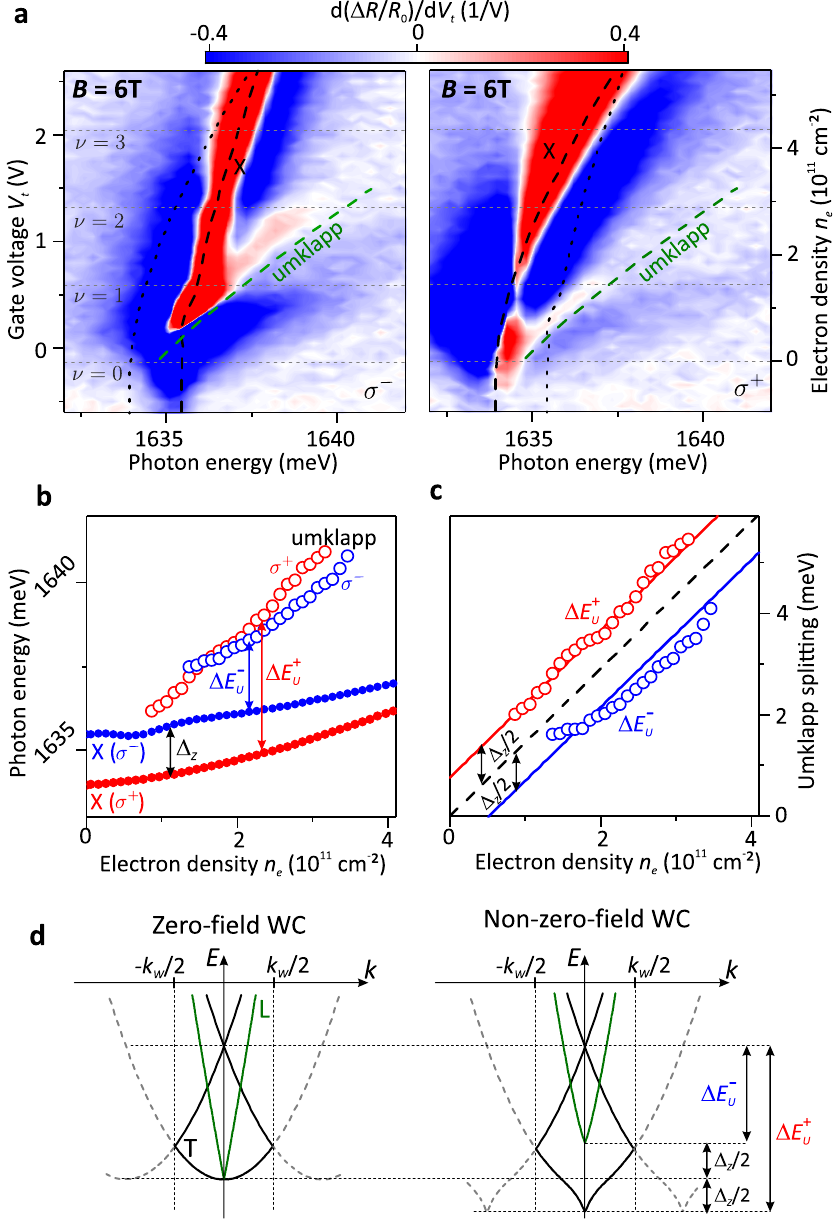}
	\caption{{\bf Enhanced Wigner crystal signatures at a magnetic field of 6~T.} ({\bf a})~Gate-voltage evolution of the derivative of the reflectance contrast spectrum with respect to $V_t$, measured at $B=6$~T in the two circular polarizations: $\sigma^-$ (left) and $\sigma^+$ (right). Grey horizontal dashed lines indicate the voltages corresponding to subsequent integer filling factors. Black dashed (dotted) lines mark the position of the co-polarized (cross-polarized) exciton resonance in each map. Green lines indicate the expected energy of the umklapp peak, which is the same for both polarizations and is computed as $E_X(\sigma^+)+\Delta_Z/2+h^2n_e/\sqrt{3}m_X$ assuming triangular electron lattice and $m_X=1.2m_e$. Here, $n_e$ denotes the electron density, $\Delta_Z$ is the exciton Zeeman splitting and $E_X(\sigma^+)$ is the energy of the lower-energy Zeeman-split exciton branch. ({\bf b})~Energies of the exciton and umklapp resonances determined as a function of $n_e$. ({\bf c})~Energy splitting between the umklapp and exciton resonances in the two circular polarizations. Solid lines mark the linear fits of the data corresponding to the exciton mass of $m_X=(1.2\pm0.2)m_e$. ({\bf d})~Schematics illustrating the influence of the magnetic field on the exciton dispersion. The $k=0$ states undergo a Zeeman splitting, which is suppressed for the finite-momentum umklapp resonance due to the existence of sizable exchange-induced splitting between longitudinal and transverse branches at $k=k_W$.}
	\label{fig:Fig2}
\end{figure}

These observations, in conjunction with recent work linking the appearance of high energy excitonic umklapp resonances to the presence of electronic charge order in an electronic Mott-insulator state~\cite{Shimazaki_arXiv_2020}, allow us to conclude that the electrons at low $n_e$ form a WC. Even though there is no moire potential in the single monolayer TMD sample we investigate, the charge order appearing in the WC state introduces a periodic potential for the excitons, which in turn leads to emergence of new bright resonances in the reduced excitonic Brillouin zone. These new transitions originate from the umklapp scattering of the dark exciton states with momentum $k=k_W$ (where $k_W$ is the WC reciprocal lattice vector), which folds these states back to the light cone where they hybridize with the $k=0$ exciton and thus acquire a finite oscillator strength. In the relevant limit of weak exciton-electron interactions, the energy of such an umklapp resonance is simply determined by the exciton kinetic energy $\Delta E_U=\hbar^2 k_W^2/2m_X$ at momentum \mbox{$k=k_W$}, where $m_X$ stands for the exciton mass. Since $k_W\sim1/a_W\sim\sqrt{n_e}$, the umklapp energy increases linearly with $n_e$ while the WC lattice constant $a_W$ is reduced, in full agreement with our experimental observations. Moreover, assuming a triangular lattice structure, the slope $\Delta E_U/n_e=h^2/\sqrt{3}m_X$ extracted from the data in Fig.~\ref{fig:Fig1}{\bf e} corresponds to $m_X=(1.1\pm0.2)m_e$ (where $m_e$ denotes a free electron mass) that is in agreement with $m_X=m_e^*+m_h^*=1.3m_e$ revealed by prior transport and ARPES experiments~\cite{Zhang2014, Larentis2018}. 

\begin{figure*}[t]
	\includegraphics{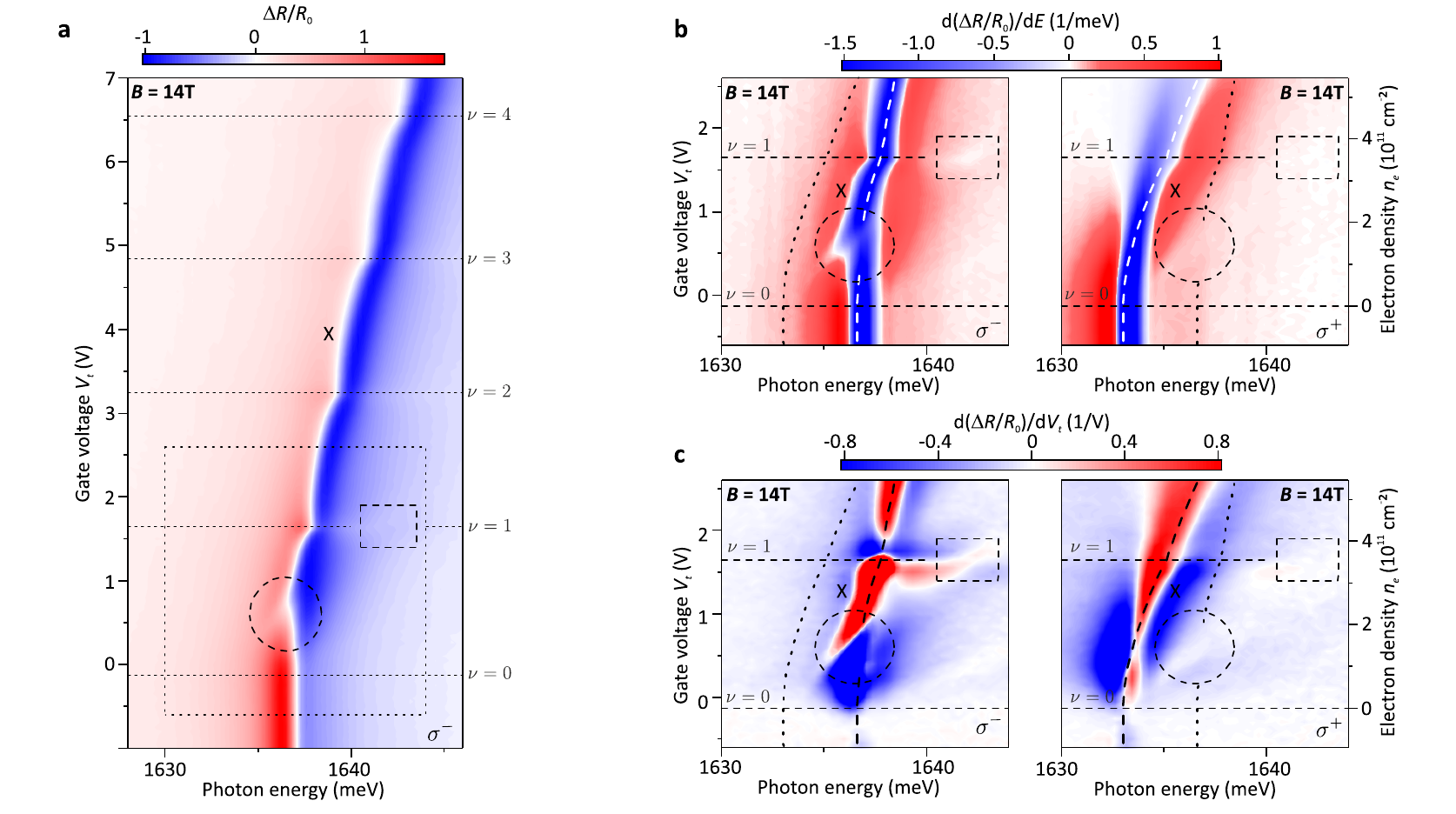}
	\caption{{\bf Interplay between Wigner crystal and integer quantum Hall state at a magnetic field of 14~T.} ({\bf a})~Gate-voltage dependence of the $\sigma^-$-polarized reflection contrast spectrum at $B=14$~T. Horizontal dashed lines mark the voltages corresponding to subsequent integer filling factors, where the exciton energy exhibits cusp-like shifts. ({\bf b-c})~Close-ups into the low-density region (marked by dotted rectangle in panel~{\bf a}) showing the gate-voltage dependence of the derivative of reflection contrast spectra with respect to the energy ({\bf b})~or the voltage ({\bf c})~in the two circular polarizations: $\sigma^-$ (left) and $\sigma^+$ (right). In each map, dashed circle indicates the area corresponding to the anticrossing between the umklapp and the exciton occurring in $\sigma^-$-polarization, while the dashed rectangle marks the region where the higher-energy resonance reappears around $\nu=1$.}
	\label{fig:Fig3}
\end{figure*}

The stability of the WC could be further enhanced upon application of the magnetic field, since confinement of the electron motion into circular orbits partially suppresses the kinetic energy. This gives rise to more pronounced optical signatures of the crystal phase, as revealed by Fig.~\ref{fig:Fig2}{\bf a} presenting gate-voltage dependence of $\mathrm{d}(\Delta R/R_0)/\mathrm{d}V_t$ acquired at $B=6$~T. Interestingly, even though the umklapp resonances in both circular polarizations are more intense than their $B=0$ counterparts, the $\sigma^-$-polarized resonance is clearly stronger than the $\sigma^+$ one. At the same time, both peaks exhibit almost identical energy for any given $n_e$, in stark contrast to the $k=0$ exciton transitions exhibiting a large valley Zeeman effect~\cite{Li2014, Srivastava_NP_2015, Aivazian_NP_2015}: \mbox{$\Delta_Z=g\mu_BB$} with $g\approx4.3$ ( Fig.~\ref{fig:Fig2}{\bf b}). This striking disparity arises from the long-range electron-hole exchange interaction~\cite{Yu2014, Qiu2015, Glazov_PRL_2014}, which strongly mixes high-momentum excitons in $K^\pm$ valleys, resulting in  formation of longitudinal- and transverse-polarized exciton branches that are split by $\Delta E_{e-h}(k)\sim|k|$, as schematically depicted in Figs~\mbox{\ref{fig:Fig1}{\bf f-h}}. Since $\Delta E_{e-h}(k_W) \gg \Delta_Z$ at $B=6$~T even for $k_W$ corresponding to $n_e \sim 1\cdot10^{11}\ \mathrm{cm}^{-2}$ (see Sec. S9 of the SI for the estimation of $\Delta E_{e-h}$), the Zeeman effect for the umklapp resonances is almost fully suppressed (Fig.~\ref{fig:Fig2}{\bf d}). Consequently, experimentally-determined splittings $\Delta E^\pm_U$ of the umklapp from the exciton in $\sigma^\pm$ polarization (Fig.~\ref{fig:Fig2}{\bf c}), despite exhibiting the same increasing slope with $n_e$ as in the zero-field case, now extrapolate to non-zero energy of $\pm\Delta_Z/2$ at $n_e=0$. These observations further support our identification of the umklapp resonance. The lower value of $\Delta E^-_U<\Delta E^+_U$ also implies stronger hybridization of the $\sigma^-$-polarized umklapp resonance with the $k=0$ excitons, which in turn explains its markedly higher intensity.

\begin{figure*}[t]
	\includegraphics{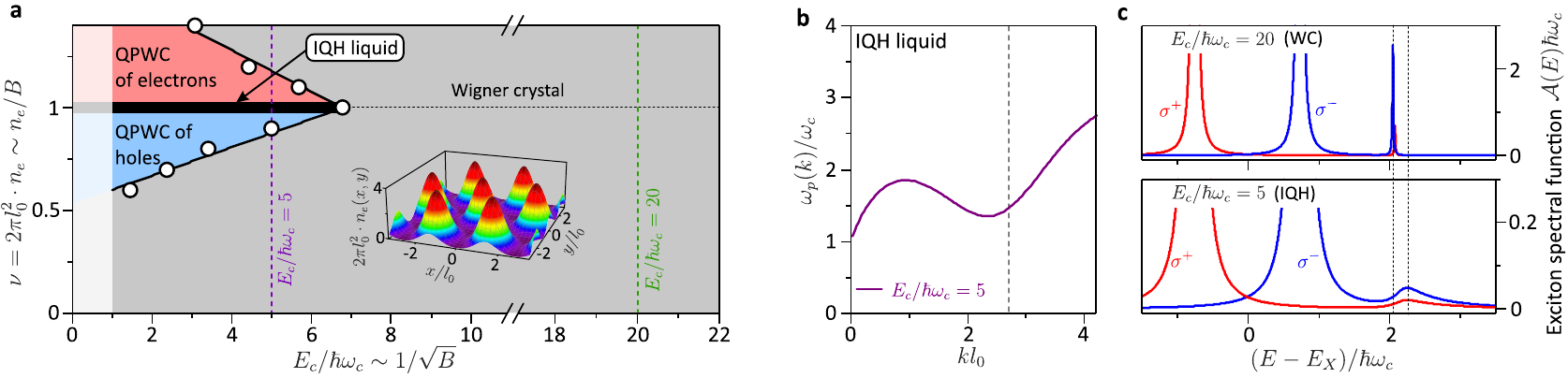}
	\caption{{\bf Theoretical analysis.} ({\bf a}) The Hartree-Fock phase diagram. We identify four phases: the liquid integer quantum Hall (IQH) state at $\nu = 1$, the conventional WC with one electron per unit cell, quasiparticle Wigner crystal (QPWC) of excess electrons for $\nu = 1+\epsilon$ and of excess holes for $\nu = 1-\epsilon$. Note the phase transition between the IQH liquid and the WC crystal at $E_c/\hbar \omega_c= 6.8$. The shaded region, $E_c\lesssim \hbar \omega_c$, labels the more conventional regime, where fractional QH states are expected to emerge. The inset shows a real-space density profile of the WC computed at $E_c/\hbar\omega_c = 10$ and $\nu = 1$. ({\bf b}) The spectrum of magnetorotons of the $\nu = 1$ liquid state for $E_c/\hbar \omega_c = 5$. This dispersion exhibits a magnetoroton minimum around the reciprocal lattice vector of the triangular WC (dashed vertical line). ({\bf c}) Comparison of the exciton spectral function at $\nu = 1$, when the electrons are in the WC (top panel) or the liquid IQH state (lower panel). The broad high-energy peak in the IQE state occurs due to exciton dressing by the soft magnetoroton mode. While the integrated spectral weight of the higher-energy peaks is similar for both the WC and the liquid states (and yields around $\sim 1\%$), the peaks are much sharper in the case of WC.}
	\label{fig:Fig4}
\end{figure*}

Combination of vanishing Zeeman-splitting of the umklapp resonance and the large valley Zeeman shift of $k=0$ excitons suggests that it is possible to bring these two transitions into resonance by tuning the $B$-field such that $\Delta_Z/2=\hbar^2k_W^2/2m_X$. Notably, at strong fields this condition could be met for $n_e$ high enough to allow for a direct observation of an anticrossing between the two resonances. This is experimentally realized at $B=14$~T, where we indeed discover such an anticrossing around $n_e=n_a\approx1.5\cdot10^{11}\ \mathrm{cm}^{-2}$ in $\sigma^-$ polarization, as shown in Fig.~\ref{fig:Fig3}. Remarkably, the umklapp resonance for $n_e$ satisfying $\Delta_Z/2 \ge \hbar^2k_W^2/2m_X$ is visible even in the gate-voltage dependence of the bare reflectance contrast spectra (Fig.~\ref{fig:Fig3}{\bf a}). In turn, differentiation of $\Delta R/R_0$ with respect the energy shows the onset of anticrossing between the umklapp and main $\sigma^-$ resonance (Fig.~\ref{fig:Fig3}{\bf b}). The latter plot also reveals a pronounced asymmetry in the intensities of the optical transitions involved in the anticrossing: at low densities $n_e\lesssim n_a$ both resonances are clearly visible, while at $n_e\gtrsim n_a$ the umklapp resonance is no longer discernible; this is independently confirmed by differentiating $\Delta R/R_0$ with respect to $n_e$, depicted in Fig.~\ref{fig:Fig3}{\bf c}. We tentatively attribute this observation to an electronic phase transition corresponding to the melting of the WC, which takes place at a relatively low filling factor ($\nu\approx0.5$) due to reduction of $E_c/\hbar \omega_c \propto 1/\sqrt{B}$ at high $B$ fields. Remarkably however, the higher-energy resonance reappears in the spectrum for $n_e$ corresponding to $\nu=1$ (see Figs~\ref{fig:Fig3}{\bf b-c}).

To provide a theoretical understanding of the experimental results at $B \neq 0$, we performed a Hartree-Fock analysis of the system. In the discussion below, we focus on understanding the main experimental observations and elucidating the difference between our system and previously studied quantum Hall WC states. One of the surprising results of this work is the observation of WC states at integer $\nu$. Majority of previous experimental investigations of WC have been done at $\nu < 1$, where macroscopic LL degeneracy can be lifted through the formation of a charge-density-wave (CDW) state~\cite{yoshioka1983ground,lam1984liquid,macdonald1984influence,cote1991collective}. An alternative, competing option for lifting the degeneracy is through fractional quantum Hall states. By contrast, integer quantum Hall (IQH) states are non-degenerate and incompressible and should be robust against the Coulomb interaction. The novelty of the TMD system is that $E_c/\hbar\omega_c \gg 1$ for $B\lesssim 10$~T: such a strong Coulomb interaction results in substantial LL mixing, leading to a transition between the liquid IQH and WC states as the ratio $E_c/\hbar \omega_c \propto 1/\sqrt{B}$ is varied. Another interesting feature of the experiment is the difference in the $\nu$-dependence of umklapp resonances for different $B$. When the magnetic field is $B \lesssim 6\,$T, we find that the umklapp resonance exists in a broad range of densities and does not exhibit any special features at integer filling factors (Fig.~\ref{fig:Fig2}). Conversely, for $B \gtrsim 10\,$T, the higher-energy resonance is observed at low $\nu$ and then around $\nu=1$, but disappears as the density is tuned either below or above (Fig.~\ref{fig:Fig3}). Earlier theoretical analysis suggested that quantum Hall crystals should not exhibit singularities at integer filling factors\cite{tevsanovic1989hall}, which is consistent with the low field results but is in contradiction with the higher field data. 

Fig.~\ref{fig:Fig4}{\bf a} summarizes our results for the Hartree-Fock phase diagram for the electron system at $B \neq 0$ as a function of the electron filling factor around $\nu=1$ and~$E_c/\hbar\omega_c$. Within this analysis, we consider four competing phases: the IQH liquid at $\nu = 1$, the WC phase (with one electron per unit cell), the quasiparticle WC (QPWC) of excess electrons for $\nu = 1+\epsilon$ and of excess holes for $\nu = 1-\epsilon$. Motivated by the large Zeeman splitting, we consider ferromagnetic states only. This mean-field approach cannot describe fractional quantum Hall states, which we expect to be important for smaller~$E_c/\hbar\omega_c$. For $\nu \gg 1$, a wealth of other phases, including different kinds of bubble states and stripes, are expected to emerge \cite{koulakov1996charge,fogler2002stripe}; however, our analysis indicates they are not relevant for the phase diagram in Fig.~\ref{fig:Fig4}{\bf a}. We find a first-order transition between the IQH and WC states at $E_c/\hbar\omega_c=6.8$. We have also performed Kallin-Halperin-type analysis of magnetoroton excitations in the IQH state and observed that full softening occurs at a much larger interaction strength $E_c/\hbar\omega_c=15.47$ (see SI), further indicating that the transition is first order. We note that mean-field approaches tend to overestimate the stability of the broken symmetry phases, and quantum fluctuations will shift the transition point towards larger values of $E_c/\hbar\omega_c$.

For $B \lesssim 6\,$T, we are deep in the WC phase ($E_c/\hbar\omega_c \gtrsim 20$), which explains the robustness of the umklapp peak observed over a broad range of $n_e$. For $B \gtrsim 10,$T we expect that at $\nu=1$ the system could already be in the IQH state. In this case, it would be the exciton dressing by the soft magnetoroton excitations, and not scattering off the WC, that gives rise to an extra peak in the exciton spectral function at an energy splitting $\sim \Delta E_U$, close to the umklapp resonance associated with the WC phase. Analysis of the exciton spectral function, including magnetoroton shake-off processes, is presented in the SI. As $n_e$ is tuned away from $\nu=1$, the IQH state transitions into QPWCs, which have different collective excitations and should have umklapp resonances at much smaller energy splittings. Furthermore, we expect QPWC phases to be much more sensitive to the disorder potential, which could lead to a strong broadening of the associated umklapp peaks. This provides a scenario for why at $B \gtrsim 10\,$T higher-energy resonances appear only at low $\nu$ (due to umklapp scattering off the WC) and around $\nu=1$ (due to magnetorotons), but not at intermediate $\nu$. We note however that for $B=14\,$T, we have $E_c/\hbar\omega_c\approx13$ at $\nu=1$, which according to the HF analysis, should still be in the WC state. Future theoretical studies beyond mean-field approximation should clarify the precise point of the IQH to the WC transition. 

The possibility to obtain a higher-energy excitonic resonance due to dressing of excitons with soft magnetoroton excitations of the $\nu=1$ IQH liquid, naturally brings up the question whether the spectral features which we identified as umklapp resonances could be considered as an unequivocal evidence for a WC phase. To clarify this point, we have carried out a calculation of the exciton spectral function at $B=0$ in the liquid phase near the critical point ($n_e \ge n_{\rm cr}$). We find that the higher-energy peak in this case (see SI) has an energy detuning, determined approximately by the energy of the roton mode, that is almost twice as large as its counterpart in the WC phase ($n_e \le n_{\rm cr}$). Since the uncertainty in the determination of the umklapp energy at $B=0$ is much smaller, our experiments clearly rule out a competing explanation based on dynamical screening of excitons by roton excitations in the liquid state. Effectively, the first order nature of the $B=0$ liquid-to-crystal phase transition taking place without substantial roton softening, ensures that the observation of an umklapp resonance at energy $\Delta E_U/n_e \simeq h^2/\sqrt{3}m_X$ provides an indisputable evidence for the WC phase.

The experiments and the theoretical analysis we detail in this article open up possibilities to study strongly correlated electrons in previously unexplored parameter regimes. An obvious extension of our work would be to combine it with transport spectroscopy~\cite{spivak2010colloquium} to measure Hall conductivity: such measurements will help to explore whether a Hall crystal~\cite{tevsanovic1989hall}, concurrently exhibiting unity Chern number and broken translational invariance, exists in the vast parameter regime that can be studied in charge-tunable monolayer TMDs. Additional insight into the dynamical properties of WCs can be obtained from optical conductivity measurements, which can elucidate the role of disorder through measurements of the pinning frequency~\cite{Williams_PRL_1991, ruzin1992pinning,zhu1994sliding,fogler2000dynamical,Ye2002,chitra2005zero, Chen_NatPhys_2006}. Our calculations show that in the absence of disorder, the generically broad asymmetric lineshapes associated with the dynamical screening of excitons by softened magnetorotons contrast sharply with the delta-function umklapp peak of a WC (Fig.~\ref{fig:Fig2}{\bf c}): consequently, we expect materials with reduced disorder to allow for an all-optical investigation of the dynamical response function of strongly interacting electrons.

Another exciting direction is the investigation of emerging spin order and its relation to the crystalline structure~\cite{bernu2001exchange}. In this context, we note that increasing resonant light intensity could be used to generate a nonequilibrium electron spin population and to study new magnetic phases that are otherwise not accessible in the electronic ground state. Further control over the ratio of Coulomb interaction to kinetic energy scales could be obtained by using more complex van der Waals heterostructures composed of bilayer semiconductors or proximal graphene layer to change the screening of interactions. On the theoretical side our work calls for further theoretical analysis of quantum Hall systems in the previously unexplored regime of ultra-strong interactions, when Coulomb interaction is much larger than the cyclotron energy. 

\section*{Acknowledgments}
C.K. and A.I. thank M. Knap for many insightful discussions. This work was supported by the Swiss National Science Foundation (SNSF) under Grant No. 200021-178909/1 and the European Research Council (ERC) Advanced Investigator Grant (POLTDES). K.W. and T.T. acknowledge support from the Elemental Strategy Initiative conducted by MEXT, Japan, A3 Foresight by JSPS and CREST (grant no. JPMJCR15F3) and JST. P.E.D, I.E., and E.D. were supported by Harvard-MIT CUA, AFOSR-MURI: Photonic Quantum Matter award FA95501610323, Harvard Quantum Initiative.

\end{document}



\title{Supplementary Information for \\ ``Observation of Wigner crystal of electrons in a monolayer semiconductor''}

\author{T. Smole\'nski}
\affiliation{\ETH}

\author{P. E. Dolgirev}
\affiliation{\HARVARD}

\author{C. Kuhlenkamp}
\affiliation{\ETH}
\affiliation{\TUM}
\affiliation{\MCQST}

\author{A. Popert}
\affiliation{\ETH}

\author{Y. Shimazaki}
\affiliation{\ETH}

\author{P.~Back}
\affiliation{\ETH}

\author{M. Kroner}
\affiliation{\ETH}

\author{K.~Watanabe}
\affiliation{\NIMSRCFM}

\author{T.~Taniguchi}
\affiliation{\NIMSICMN}

\author{I. Esterlis}
\affiliation{\HARVARD}

\author{E. Demler}
\affiliation{\HARVARD}

\author{A. Imamo\u{g}lu}
\affiliation{\ETH}

\maketitle

\section{Device and experimental setup}
\label{sec:sample_setup}

\begin{figure*}
\includegraphics{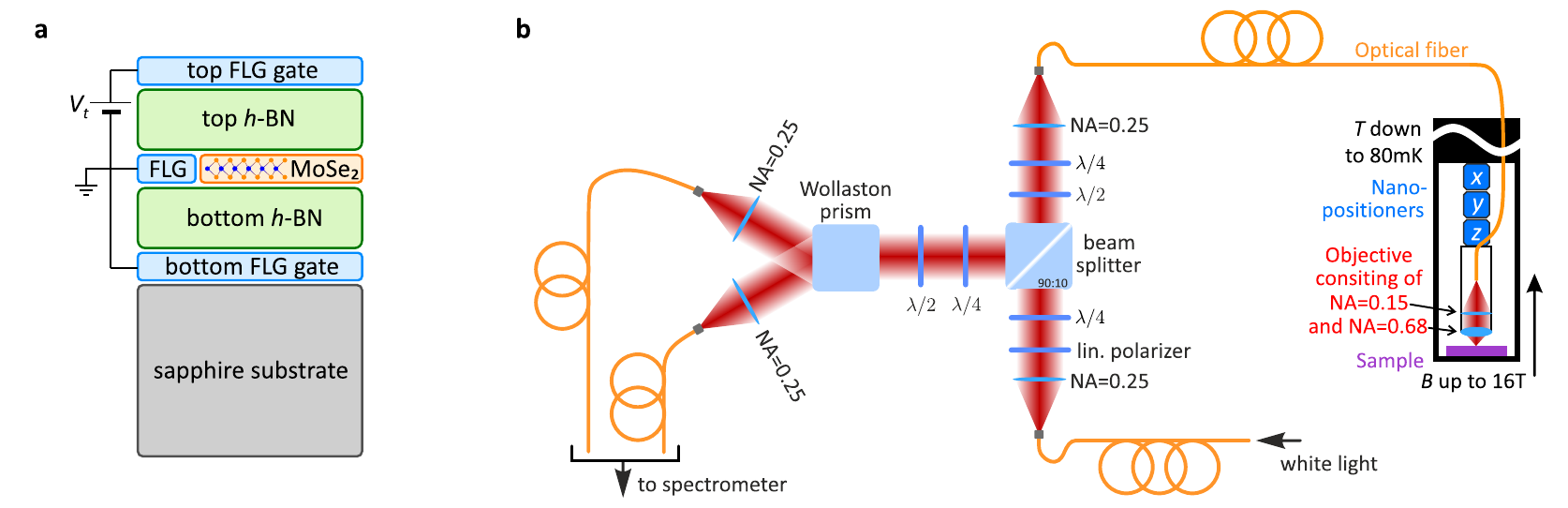}
\caption{{\bf Device structure and the experimental setup.} ({\bf a})~Cartoon displaying the structure of the device investigated in the main text. ({\bf b})~Simplified schematic of the experimental setup utilized for magneto-optical measurements.}\label{fig:S1}
\end{figure*}

The structure of the device studied in the main text is schematically depicted in Fig.~\ref{fig:S1}{\bf a}. The device consisted of a charge-tunable MoSe$_2$ monolayer that was electrically contacted with a few-layer graphene (FLG) flake, encapsulated between two hBN flakes, and finally embedded between the top and bottom FLG graphene gates. The thicknesses of hBN layers -- being particularly important for modelling of the spectral profiles of exctionic resonances -- were determined by means of atomic force microscopy (AFM) to be $t_t=(74\pm5)$~nm and $t_b=(91\pm5)$~nm for the top and bottom layers, respectively. All of the flakes were first mechanically exfoliated either from synthetic (HQ Graphene MoSe$_2$, NIMS hBN) and natural (graphene) bulk crystals onto Si/SiO$_2$ substrates using a backgrinding tape (Ultron Systems). The actual heterostructure was assembled with a standard, dry-transfer technique~\cite{Zomer_APL_2014}, in which the flakes were sequentially picked up and stacked together with sub-micron spatial precision using of a hemispherical, polydimethylsiloxane (PDMS) stamp covered with a thin layer of polycarbonate (PC). A complete stack was then released on a transparent, sapphire substrate, and the residual PC layer was subsequently dissolved in chloroform. Finally, the FLG graphene flakes were contacted with Ti/Au electrodes that were fabricated using standard electron-beam lithography followed by metal evaporation. During all of the reported experiments, the electron density $n_e$ in the MoSe$_2$ monolayer was controlled by applying a voltage $V_t$ to the top gate, while keeping the MoSe$_2$ and the other gate grounded. Notably, the sample exhibited virtually no electrical hysteresis when the gate voltage was swept upwards or downwards, as the difference in values of $V_t$ being required to reach a given density in these two cases was not exceeding 0.1~V. Nonetheless, in order to avoid this small uncertainty, in all of our experiments the gate was always ramped in the same direction (from negative to positive values).

Figure~\ref{fig:S1}{\bf b} displays a simplified schematic of the experimental setup used in our measurements. The device was mounted inside a dilution refrigerator allowing to reach a base temperature of 80~mK. The refrigerator was immersed in a liquid He bath cryostat equipped with a superconducting solenoid producing a magnetic field of up to 16~T in the direction perpendicular to the sample surface. The resonant-reflection of the device was measured with the use of a broadband light emitting diode (LED) with center wavelength of 760~nm and linewidth of 20~nm. The LED light was first arbitrarily polarized using a linear polarizer as well as a set of $\lambda/2$ and $\lambda/4$ wave-plates. Then it was transmitted to the sample in a single-mode fiber. The fiber was coupled inside the refrigerator to a confocal microscope objective consisting of two aspheric lenses of numerical apertures NA = 0.15 (on the fiber side) and NA = 0.68 (on the sample side) allowing to focus the light to a diffraction-limited spot of $\sim0.5\ \mu$m diameter. The integrated power of the light was kept on a low level of about 15~nW to avoid heating of the sample. The objective was mounted on $x$--$y$--$z$ piezo-electric attocube stages, which permitted to select a suitable spot on the sample surface with submicron precision. The light reflected off the sample was collected by the same fiber and then analyzed in the detection path consisting of $\lambda/2$ and $\lambda/4$ wave-plates as well as a Wollaston prism, which allowed to separate the beam into two components corresponding to $\sigma^+$- and $\sigma^-$-polarized response of the sample. Both of these components were then spectrally-resolved by the same 0.5~m spectrometer and recorded in parallel using a nitrogen-cooled CCD camera. In order to compensate for the Faraday rotation in the fiber, the polarization settings were readjusted based on the resonances observed in the reflectance spectra after each change of the magnetic field. Moreover, in order to enhance the fidelity of the detected polarization, the incoming white-light was always co-polarized to the currently analyzed $\sigma^\pm$-polarized response of the sample.

\section{Details of normalization and differentiation of the reflectance spectra}
\label{sec:norm_and_diff}

\begin{figure*}[b]
\includegraphics{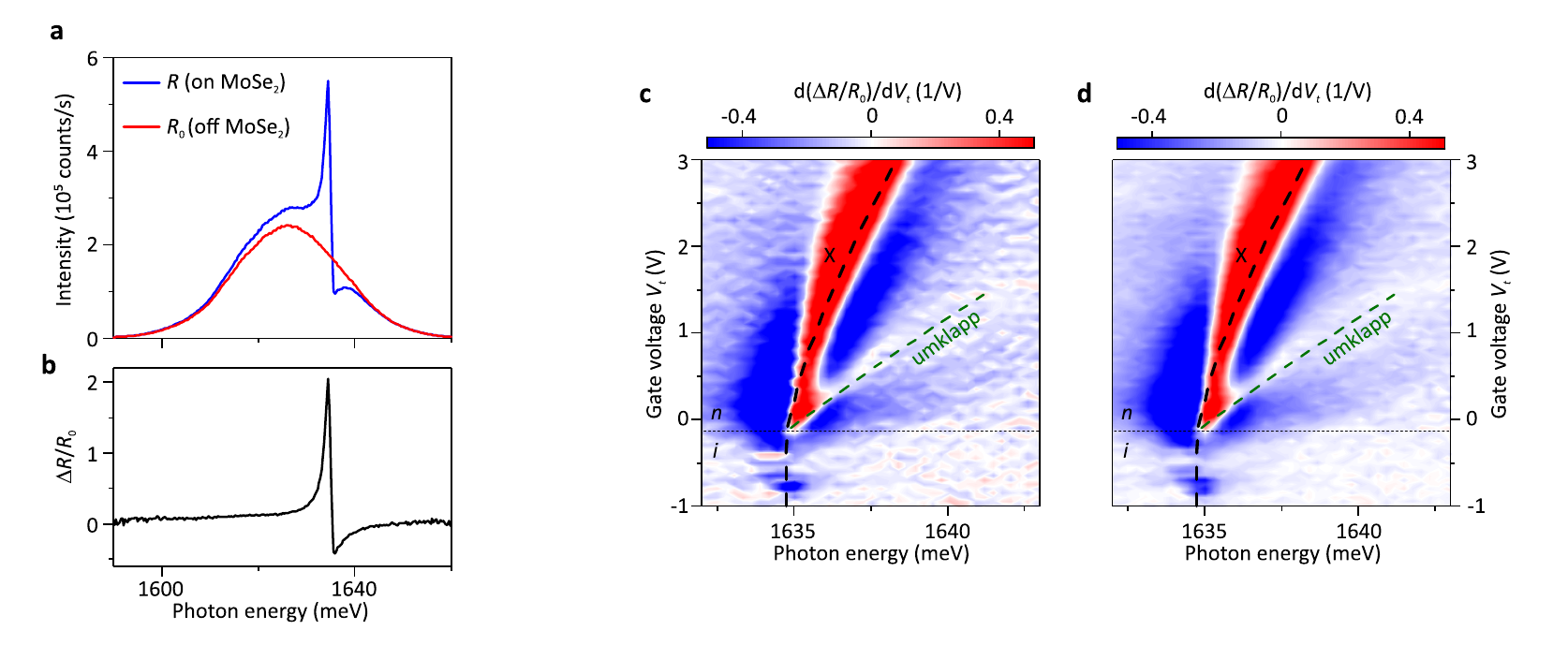}
\caption{{\bf Normalization and differentiation of the reflectance data.} ({\bf a})~Reflectance spectra acquired for the main device at two different spots: one in the MoSe$_2$ monolayer region (blue) and one off the MoSe$_2$ monolayer (red). Both spectra were obtained at charge neutrality ($V_t=-1$~V) and in the absence of the magnetic field. ({\bf b}) The reflectance contrast spectrum $R_c\equiv\Delta R/R_0$ determined based on the two spectra from panel {\bf a}. ({\bf c}-{\bf d})~Color-scale plots showing zero-field gate-voltage evolutions of the derivative of reflectance contrast $R'_c=\mathrm{d}R_c/\mathrm{d}V_t$ with respect to the $V_t$ (dashed lines mark the exciton and umklapp energies). The left panel presents a derivative evaluated numerically using standard, symmetric difference quotient method as $R'_c(V_{t,n})=[R_c(V_{t,n+1})-R_c(V_{t,n-1})]/[V_{t,n+1}-V_{t,n-1}]$. The right panel shows a derivative of the same data obtained using the other method, in which $\mathrm{d}R_c/\mathrm{d}V_t$ is computed as a difference quotient between symmetric data points separated not by two, but by four gate-voltage steps, i.e.,  $R'_c(V_{t,n})=[R_c(V_{t,n+2})-R_c(V_{t,n-2})]/[V_{t,n+2}-V_{t,n-2}]$. This method was used in the main text to plot the gate-voltage derivatives in Figs~1--3.} \label{fig:S2}
\end{figure*}

As stated in the main text, in order to extract the reflectance contrast $R_c$ from the measured white-light reflectance spectrum $R$ of the MoSe$_2$ monolayer, we acquire a reference spectrum $R_0$ at a different spot in the sample region featuring all the layers (i.e., hBN, graphene gates) except for the MoSe$_2$. To minimize systematic errors stemming from sample inhomogeneities or defocusing when travelling along the sample surface, the reference spot is selected as close as possible to the investigated MoSe$_2$ spot. The example pair of such $R$ and $R_0$ spectra is shown in Fig.~\ref{fig:S2}{\bf a}. Based on this data we determine the reflectance contrast as $R_c\equiv\Delta R/R_0=(R-R_0)/R_0$ (as displayed in Fig.~\ref{fig:S2}{\bf b}) as well as its dependence on the $V_t$ by applying the same procedure to the reflectance spectra $R(V_t)$ measured for different gate voltages.

To evaluate a derivative $R_c'(E)=\mathrm{d}R_c/\mathrm{d}E$ of the reflectance contrast with respect to the energy (which is plotted in Fig.~3{\bf b} in the main text), we apply a standard symmetric difference quotient method, which approximates a numerical derivative as a slope of the line connecting the two neighbouring data points $R'_c(E_n)=[R_c(E_{n+1})-R_c(E_{n-1})]/[E_{n+1}-E_{n-1}]$, where index $n$ labels subsequent points. In principle, similar method could be also applied for computing a derivative of $R_c$ versus gate voltage as $R'_c(V_{t,n})=[R_c(V_{t,n+1})-R_c(V_{t,n-1})]/[V_{t,n+1}-V_{t,n-1}]$. However, due to the small gate-voltage step used in our measurements, the signal differentiated in such a way is relatively noisy, as shown in Fig.~\ref{fig:S2}{\bf c} for the case of zero-field reflectance contrast spectra. For this reason, in Figs 1{\bf d}, 2{\bf a}, and 3{\bf c} in the main text we employ a slightly different procedure, in which $R_c'(V_t)$ is determined as a slope connecting not neighbouring, but next-neighbouring data points: $R'_c(V_{t,n})=[R_c(V_{t,n+2})-R_c(V_{t,n-2})]/[V_{t,n+2}-V_{t,n-2}]$. This procedure effectively reduces the noise (see Fig.~\ref{fig:S2}{\bf d}) by suppressing the contribution to $R_c'(V_t)$ from abrupt changes of $R_c$ occurring within $\Delta V_t<0.2$~V, which are irrelevant for the effects investigated in our work.

\section{Determination of energies of the exciton and umklapp resonances} 

\begin{figure*}[b]
\includegraphics{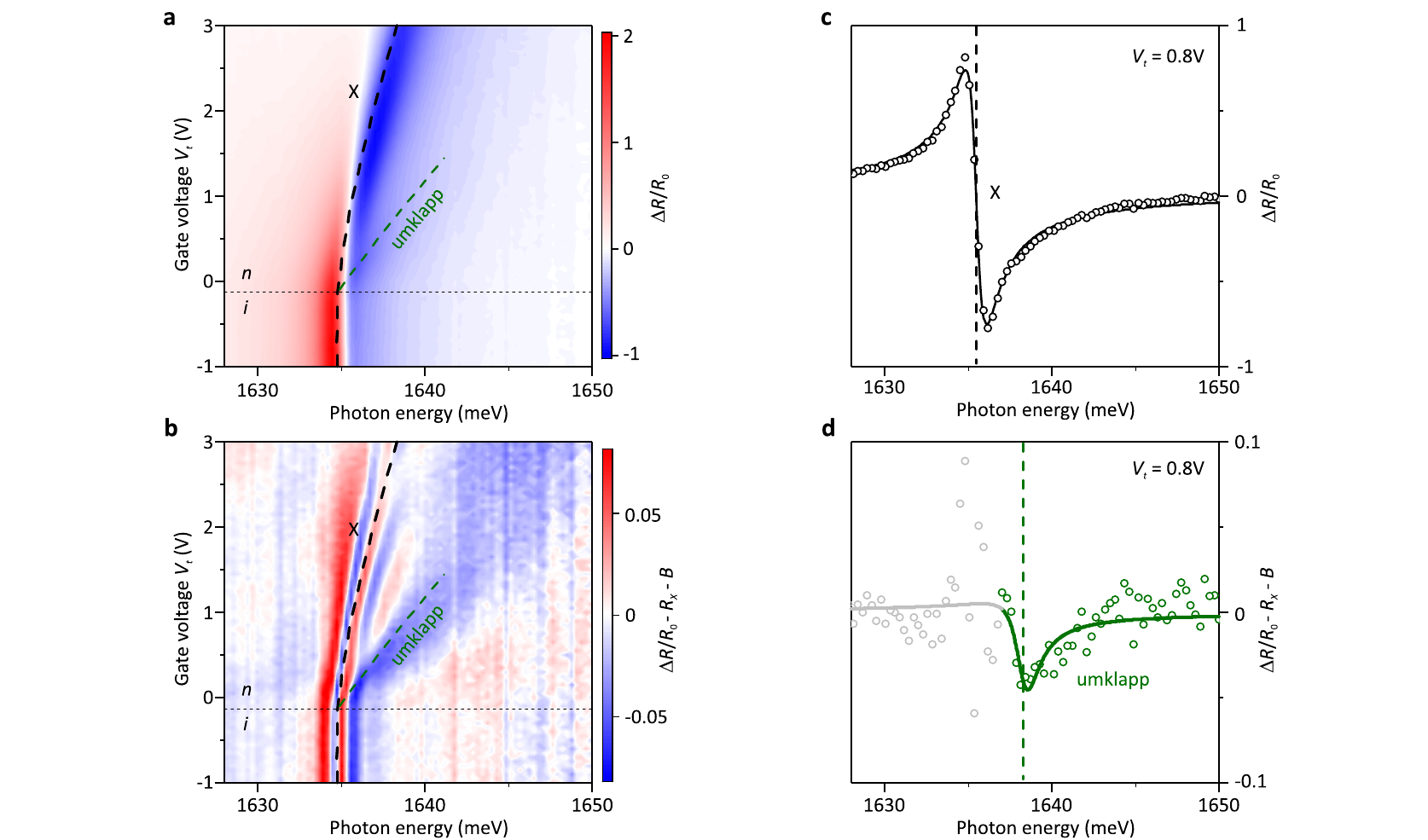}
\caption{{\bf Fitting the exciton and umklapp peaks with dispersive Lorentzian spectral profiles.} ({\bf a}) Color-scale map showing gate-voltage evolution of the reflectance contrast $R_c$ spectra measured for the main device, at zero magnetic field, for low electron doping densities, and in the spectral vicinity of the exciton peak. ({\bf b}) Similar map presenting evolution of the reflectance contrast $R_c$ upon subtraction of the fitted exciton spectral profile $R_X$ as well as a smooth gaussian-like background~$B$ (independent of the voltage). The dashed lines in both panels indicate the energies of the exciton and umklapp peaks. ({\bf c}-{\bf d})~Cross-sections through the maps in panels ({\bf a}-{\bf b}) at $V_t=0.8$~V showing, respectively, bare and background-corrected reflectance contrast spectra. The solid lines indicate the fits to the experimental data with dispersive Lorentzian spectral profiles, based on which we determined the exciton and umklapp energies (marked by vertical dashed lines). In case the umklapp peak the fitting was carried out only in the energy region covered by the data points shown in green, in order to avoid spurious contribution originating from the residual of the exciton resonance fitting.} \label{fig:S3}
\end{figure*}

Due to the presence of multiple reflections at the interfaces between different layers in our device, the light reflected off the MoSe$_2$ monolayer interferes with the background signal, which sizably alters the lineshape of the excitonic resonances observed in the spectra. In order to account for this effect, we describe each of these resonances using an effective dispersive Lorentzian spectral profile of the form:~\cite{Smolenski_PRL_2019, Shimazaki_Nature_2020}
\begin{equation}
\label{eq:fitting_formula}
R_c(E)=A\cos(\varphi)\frac{\gamma/2}{(E-E_0)^2+\gamma^2/4}+A\sin(\varphi)\frac{E_0-E}{(E-E_0)^2+\gamma^2/4}+C,
\end{equation}
where $E$ denotes the photon energy, $C$ represents a flat background, $A>0$, $E_0$, and $\gamma$ correspond, respectively, to the amplitude, energy, and linewidth of the resonance, while $\varphi$ stands for interference-induced phase-shift, which depends both on the energy and the amplitude of the resonance.

To extract the energies of the exciton and umklapp peaks from a given reflectance contrast spectrum, we first fit the spectral profile $R_X(E)$ of the exciton peak with the aforementioned dispersive Lorentzian formula leaving the umklapp peak aside, which is justified owing to its very small oscillator strength. The result of such a fit performed for an example zero-field spectrum measured at $V_t=0.8$~V is presented in Fig.~\ref{fig:S3}{\bf c}. To determine the umklapp position, we subtract the fitted lineshape $R_X(E)$ from the original data $R_c(E)$ and repeat this for each gate voltage, thus obtaining a voltage evolution of the corrected spectra $R_c(E, V_t)-R_X(E, V_t)$. At this stage, to account for the presence of a residual smooth background in the data, we fit one of the corrected spectra obtained at the charge neutral region (where we do not expect to see an umklapp signature) with a phenomenological, gaussian formula $B(E)=A_b\exp[-(E-E_b)^2/2S_b^2]+C_b$. This fixed background is then subtracted from all the spectra measured at different voltages. The resulting gate-voltage dependence of $R_c(E, V_t)-R_X(E, V_t)-B(E)$ is displayed in Fig.~\ref{fig:S3}{\bf b} together with the corresponding raw reflectance contrast $R_c(E, V_t)$ data in Fig.~\ref{fig:S3}{\bf a} (for comparison). As seen, the above approach greatly reduces the non-linear background around the umklapp peak, which makes it possible to fit its spectral profile. To this end, in order to avoid the spurious contribution from residuals of the exciton fitting, we truncate the fitting range on the red side of the umklapp resonance to the energies $E>E_X+\gamma_X$, where $E_X$ and $\gamma_X$ represent, respectively, the fitted energy and linewidth of the exciton resonance. Moreover, bearing in mind relatively low intensity of the umklapp resonance, we also reduce the number of independent parameters in the dispersive Lorentzian formula~(\ref{eq:fitting_formula}). Specifically, we set $C=0$ (since the background was already subtracted) and fix the phase of the umklapp peak to $\varphi=\varphi_0$, where $\varphi_0=2.5$~rad represents the zero-oscillator-strength limit of the resonance phase at the umklapp energy in the case of our heterostructure. The latter value is determined based on the transfer-matrix simulations~\cite{Back_PRL_2018} of our device reflectivity spectrum (assuming hBN thicknesses mentioned in Sec.~\ref{sec:sample_setup} as well as hBN refractive index of 2.10 taken from~\cite{Lee_PSSB_2019}), the validity of which is independently confirmed by ensuring that those simulations properly describe the gate-voltage evolution of the lineshapes of the main resonance in the optical spectrum (i.e., attractive and repulsive polarons). Under the above assumptions, we fit the umklapp peak in the corrected spectrum (as shown in Fig.\ref{fig:S3}{\bf d} for an example case of $V_t=0.8$~V), which finally allows us to extract its energy as a function of the gate voltage, which is plotted in Figs 1{\bf e} and 2{\bf b},{\bf c} in the main text for the cases of $B=0$ and $B=6$~T, respectively. We stress that due to large number of steps involved in the above approach, it may be fraught with a systematic error originating, e.g., from the uncertainty of assumed umklapp peak phase $\varphi_0$. This may in turn lead to a systematic error of the exciton mass determined based on the slope of the umklapp energy increase with the gate voltage. We account for this effect in the main text by assuming the mass uncertainty to be larger than its statistical uncertainty determined by the spread of the data points in Figs 1{\bf e} and 2{\bf b},{\bf c}.

\section{Calibration of the electron density dependence on the gate voltage} 

\begin{figure*}[t]
\includegraphics{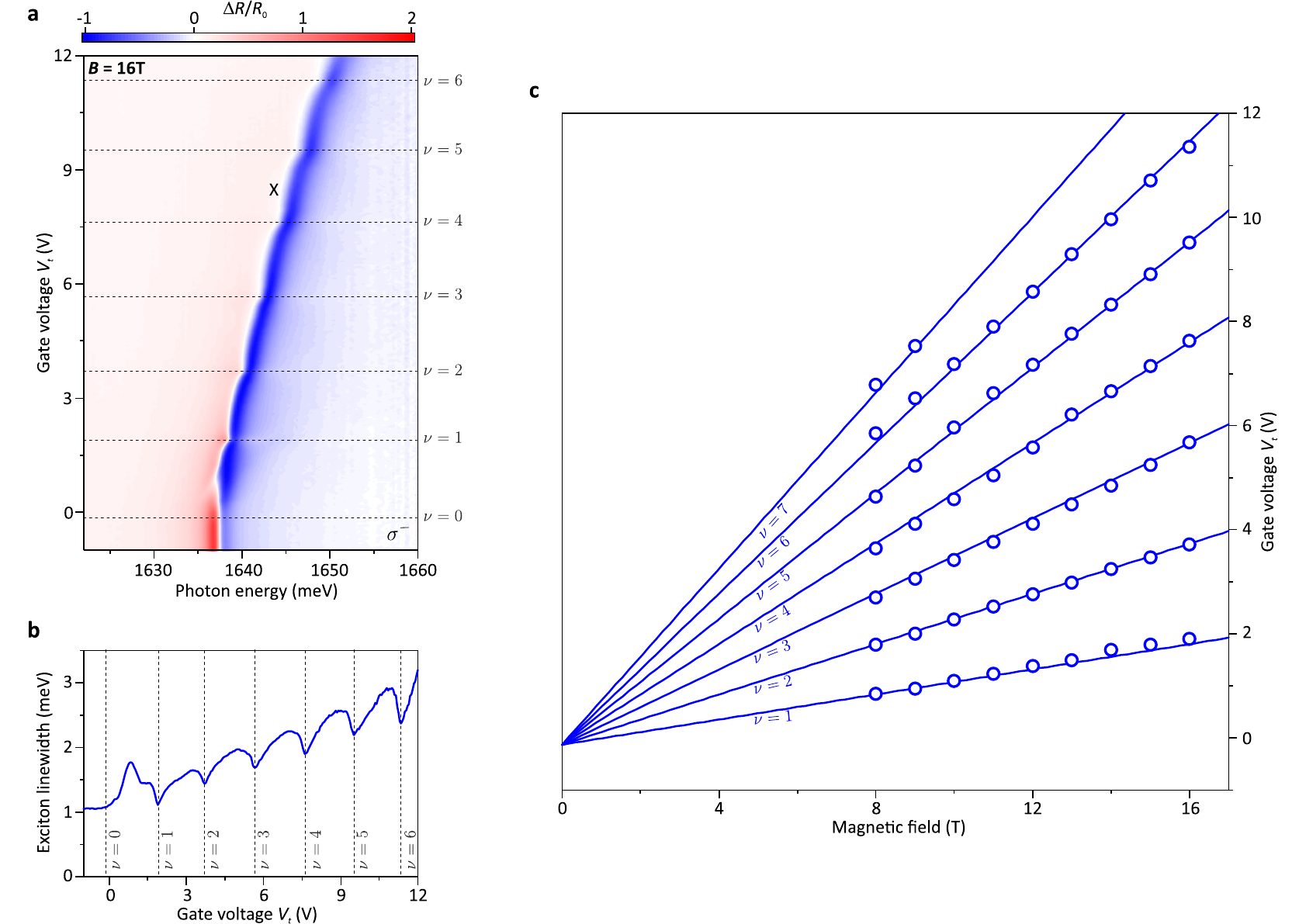}
\caption{{\bf Calibration of the electron density based on the Shubnikov-de Haas oscillations in the exciton linewidth.} ({\bf a}) Color-scale map presenting $\sigma^-$-polarized reflectance contrast spectra measured for the main device as a function of the gate-voltage at $B=16$~T. ({\bf b}) Gate-voltage dependence of the exciton linewidth extracted from the data in panel ({\bf a}) by fitting the exciton resonance with dispersive Lorentzian spectral profile. Dashed lines in both panels indicate the positions of integer filling factors. ({\bf c}) Gate voltages $V_t(\nu, B)$ corresponding to the positions of the exciton linewidth minima extracted from the reflectance measurements carried out at different magnetic fields. All of the presented data points were obtained in the regime where the Fermi level does not exceed the valley Zeeman splitting in the conduction band. Solid lines represent the fit of the data points with a set of linear dependencies corresponding to subsequent integer filling factors, which form a LL fan chart.} \label{fig:S4}
\end{figure*}

Given that the energy splitting between the umklapp and exciton transitions is governed by the Wigner crystal lattice constant that is evaluated based on the electron density $n_e$, it is essential for the validity of our analysis to precisely establish a dependence of $n_e$ on the gate voltage $V_t$. In principle, this might be done by modelling the device as a parallel-plate capacitor~\cite{Back_PRL_2018, Smolenski_PRL_2019} with the geometrical capacitance per unit area $C_\mathrm{geom}=\epsilon_0\epsilon^\perp_\mathrm{hBN}/t_t$ defined by the top hBN thickness $t_t$ and its static dielectric constant $\epsilon^\perp_\mathrm{hBN}$. This procedure, however, may be fraught with a systematic error stemming from sizable uncertainty of the $\epsilon^\perp_\mathrm{hBN}$ constant. 

To avoid the above difficulties, we utilize a more accurate approach, in which $n_e(V_t)$ is obtained based on the Shubnikov-de Haas (SdH) oscillations of the exciton transition at high magnetic fields. As we demonstrated in our previous work~\cite{Smolenski_PRL_2019}, under such conditions the linewidth $\gamma_X$ of the exciton peak in $\sigma^-$ polarization (in case of the electron doping) exhibits sharp minima each time the Landau-level (LL) filling factor $\nu$ takes on an integer value, as seen in Figs~\ref{fig:S4}{\bf a},{\bf b} for an example set of data taken at $B=16$~T. This allows us to extract the voltages $V_t(\nu,B)$ corresponding to subsequent integer $\nu$ by fitting the vicinity of each minimum in $\gamma_X(V_t)$ dependence with a phenomenological, gaussian profile. Bearing in mind that our Wigner crystal investigation from the main text is mostly focused on the low-doping density limit, we simplify the current analysis by restricting it to a regime, in which the Fermi level does not exceed the valley Zeeman splitting of the conduction band, and hence where the electrons occupy the states in a single, $K^+$ valley. Fig.~\ref{fig:S4}{\bf c} displays the values of $V_t(\nu,B)$ obtained in the above-described way for all magnetic fields $B\ge8$~T at which we observe resolvable SdH oscillations in the exciton linewidth. As expected, the voltages form a characteristic LL fan chart exhibiting linear increase with both $\nu$ and $B$, as confirmed by their perfect agreement with a set of linear dependencies of the form $V_t(\nu,B)=V_0+\delta\nu B$. Importantly, all of these dependencies have just two common fitting parameters: $V_0=-0.13$~V being the voltage that corresponds to the onset of filling the conduction band with electrons, and $\delta=0.12\ \mathrm{V/T}$ representing a change of the gate voltage that is required to fill a single LL in a unit magnetic field. Taking into account that the investigated LLs are fully spin- and valley-polarized, each of the above-introduced linear dependencies corresponds to an electron density of $n_e(\nu,B)=\nu\cdot eB/h$. This finally allows us to extract the $n_e(V_t)$ by linearly-interpolating between subsquent integer filling factors:
\begin{equation}
n_e(V_t>V_0)=\frac{e}{h\delta}(V_t-V_0)=2.00\cdot10^{11}\ \frac{\mathrm{cm}^{-2}}{\mathrm{V}}\cdot(V_t+0.13\ \mathrm{V}).
\end{equation}
This expression was used to calibrate the electron density in all of the figures in the main text. It is noteworthy that the capacitance $C=e\Delta n_e/\Delta V_t=e^2/h\delta=0.32\ \mathrm{nF/mm^2}$ of the device evaluated using the above $n_e(V_t)$ dependence agrees well with that obtained within the parallel plate approximation $C_\mathrm{geom}=\epsilon_0\epsilon^\perp_\mathrm{hBN}/t_t=(0.42\pm0.10)\ \mathrm{nF/mm^2}$ for $t_t=(74\pm5)$~nm and $\epsilon^\perp_\mathrm{hBN}=3.5\pm0.5$ estimated based on the values reported in several previous works~\cite{Kim_ACSNano_2012,Laturia_2DMA_2018}. 

\section{Comparison of the Wigner crystal signatures at different temperatures}

\begin{figure*}[h]
\includegraphics{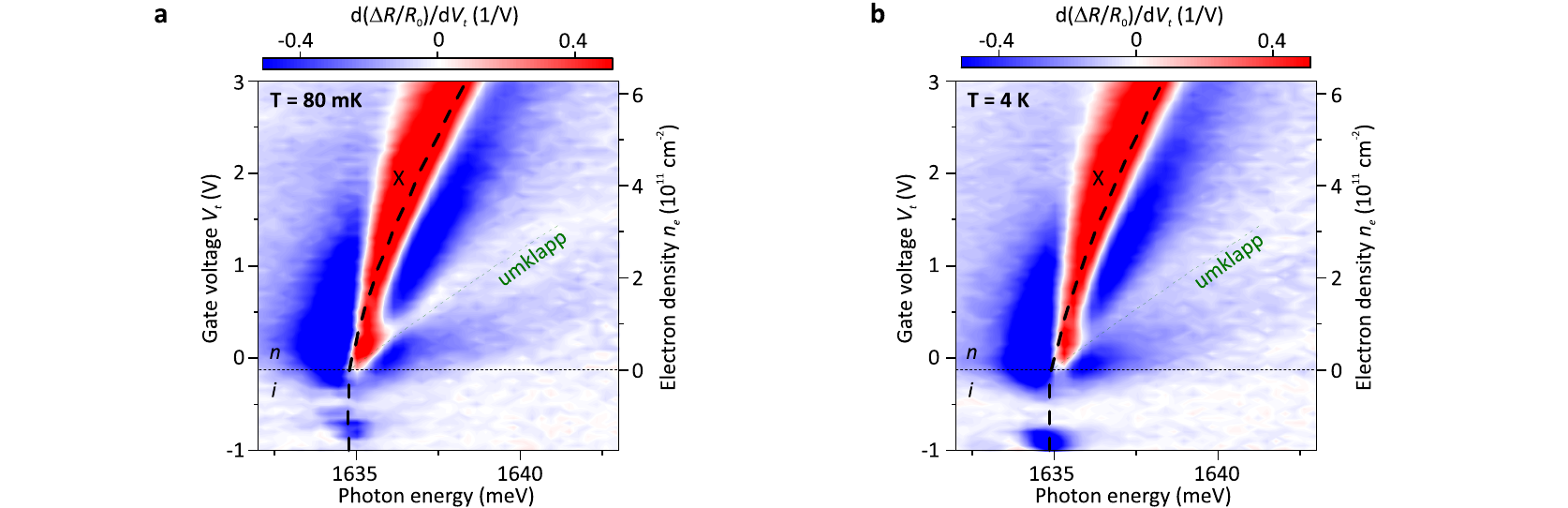}
\caption{{\bf Comparison of the zero-field Wigner crystal signatures at different temperatures.} ({\bf a}, {\bf b}) Color-scale maps showing zero-field gate-voltage evolution of the derivative of reflectance contrast spectra with respect to $V_t$ measured for the main device at two different temperatures: $T=80$~mK ({\bf a}) and $T=4$~K ({\bf b}). Black dashed lines mark the fitted energy $E_X$ of the exciton resonance, while the green lines indicate the expected position of the umklapp peak $E_X+\Delta E_U$ for $\Delta E_U=h^2 n_e/\sqrt{3}m_X$ corresponding to a triangular Wigner crystal and $m_X=1.1m_e$.} \label{fig:S5}
\end{figure*}

In order to examine the robustness of the electronic Wigner crystal against the temperature changes, we repeated the reflectance contrast measurements from the main text also at an elevated temperature of $T=4$~K. These experiments were carried after removing the $^3$He--$^4$He mixture from the mixing chamber of the dilution refrigerator and filling its vacuum can with about 1~mbar of $^4$He exchange gas to facilitate the heat exchange between the sample and liquid helium bath inside the cryostat. Fig.~\ref{fig:S5} displays the comparison between the zero-magnetic-field gate-voltage evolutions of the reflectance contrast derivative $\mathrm{d}(\Delta R/R_0)/\mathrm{d}V_t$ acquired at $T=80$~mK and $T=4$~K. Strikingly, the umklapp signature, despite being still visible at the same energies, becomes sizably weaker upon rising the temperature. This is in stark contrast to the case of the main exciton peak, the amplitude of which remains almost not affected by the temperature change. This observation directly demonstrates that while melting temperature of the Wigner crystal exceeds 4~K, the enhanced thermal fluctuations of the electrons suppress the intensity of the exciton umklapp scattering off the electronic crystal at higher $T$.

\section{Reproducibility of the results on a different device} 

As stated in the main text, we observed similar signatures of the electronic Wigner crystal also for a second device, which was fabricated using the same technique as the main one. In general, this device featured a more complex structure, but in the present study we focused exclusively on the region where the layer arrangement was similar to the case of the main device, and consisted of a dual-graphene-gated MoSe$_2$ monolayer that was fully-encapsulated between the two hBN layers. The thicknesses of those hBN flakes (measured with AFM) were slightly smaller and yielded $t_t=(41\pm5)$~nm and $t_b=(50\pm5)$~nm, respectively, for the top and bottom layers. Unlike the case of the main sample, the optical measurements on the second one were not carried out in a dilution refrigerator, but in a standard dipstick filled with helium exchange gas and immersed in a liquid helium bath cryostat, allowing to reach the sample temperature of $T=4$~K. The electron density $n_e$ was controlled by applying a top gate voltage $V_t$ while keeping the back gate and the MoSe$_2$ monolayer grounded. 

\begin{figure*}[t]
\includegraphics{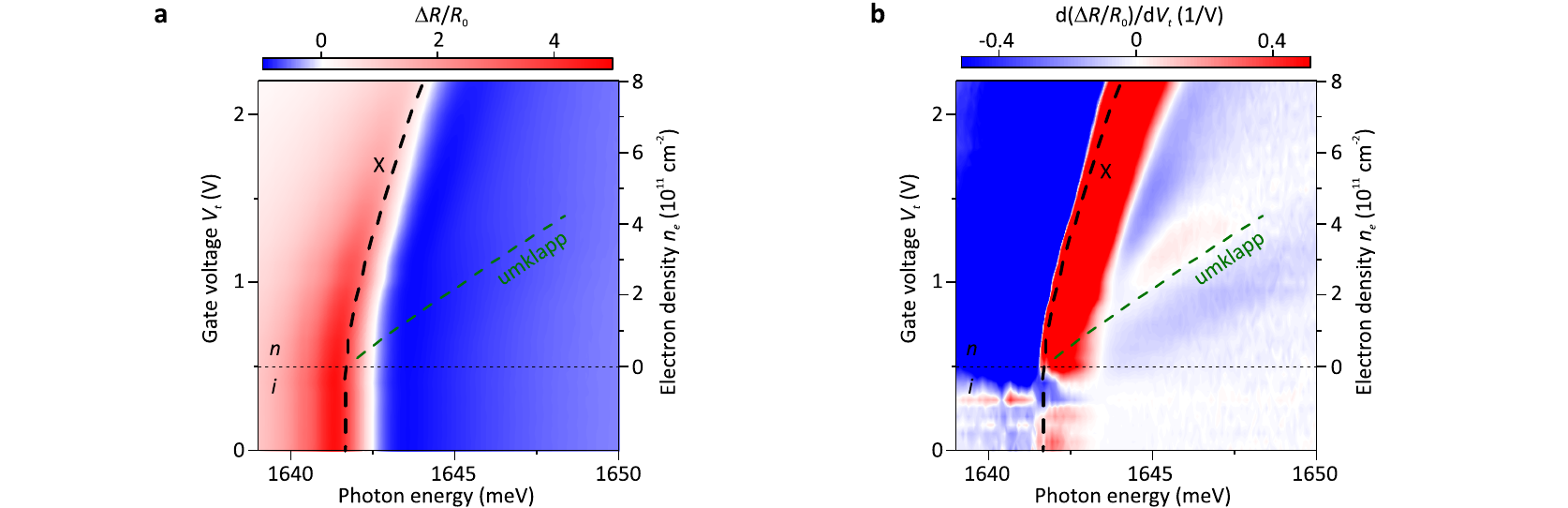}
\caption{{\bf Observation of zero-magnetic-field Wigner crystal signatures for the second device.} ({\bf a}) Color-scale map showing reflectance contrast spectra measured as a function of the top gate voltage $V_t$ for the second device. The data were acquired at $T=4$~K and in the absence of the magnetic field. ({\bf b}) Gate-voltage evolution of the derivative of the spectra from panel ({\bf a}) with respect to $V_t$. Black dashed lines in both panels indicate the energy of the exciton peak $E_X$ obtained by fitting its spectral profile with dispersive Lorentzian lineshape. Green lines mark the expected position $E_X+\Delta E_U$ of the umklapp peak, where $\Delta E_U=h^2 n_e/\sqrt{3}m_X$ is computed under assumption of triangular Wigner crystal and for the value of exciton mass $m_X=1.3m_e$.}\label{fig:S6}  
\end{figure*}

Fig.~\ref{fig:S6}{\bf a} presents a gate-voltage evolution of the reflectance contrast spectra measured for the second sample at $B=0$ in the spectral vicinity of the exciton resonance. The corresponding voltage dependence of the derivative of $\Delta R/R_0$ with respect to $V_t$ is shown in Fig.~\ref{fig:S6}{\bf b}. The differentiated data clearly demonstrate the presence of an umklapp peak on the high-energy side of the exciton, which behaves in the same way as in the case of the main device: its detuning from the main exciton peak increases for larger electron densities, it appears to merge with the exciton in the limit of $n_e=0$, and becomes indiscernible at $n_e\gtrsim4\cdot10^{11}\ \mathrm{cm}^{-2}$. Owing to the elevated temperature of the sample, the umklapp resonance in the present experiments exhibits significantly lower intensity, which makes it difficult to directly fit its spectral profile using the same technique as for the main device. Instead, we attempt to estimate the umklapp spectral position as $E_X+\Delta E_U$, where $E_X$ represents the exciton energy extracted from dispersive Lorentzian fit, while the $\Delta E_U=h^2 n_e/\sqrt{3}m_X$ denotes the exciton-umklapp splitting that is computed under assumption of the triangular Wigner crystal geometry and for the value of the exciton mass $m_X=m_e^*+m_h^*=1.3m_e$ obtained from previous experiments on MoSe$_2$ monolayers~\cite{Zhang_NatNano_2014, Larentis_PRB_2018, Goryca_NatCommun_2019}. The electron density $n_e$ is in turn determined within a parallel-plate capacitor approximation as $n_e(V_t)=(V_t-V_0)\cdot C_\mathrm{geom}/e$, where $C_\mathrm{geom}=\epsilon_0\epsilon^\perp_\mathrm{hBN}/t_t=(0.76\pm0.22)\ \mathrm{nF/mm^2}$ stands for the geometrical capacitance of the device, while $V_0=0.5\ \mathrm{V}$ is the voltage corresponding to the onset of filling the conduction band with electrons (which is extracted from the reflectance data as $V_t$ at which the main exciton resonance starts to blueshift). Remarkably, the umklapp energy estimated in the above-defined way (marked with a green dashed line in Figs~\ref{fig:S6}{\bf a-b}) remains in a good agreement with the actual peak position. This finding further supports the identification of the umklapp peak and confirms that the investigated electronic Wigner crystallization is not specific to a single device, but is a general characteristic of TMD monolayers.

\section{Instabilities of the IQH liquid state at $\nu = 1$} 
We consider the microscopic Hamiltonian:
\begin{align}
    \hat{H}_{ee} & = \int d {\bf r} \, \hat{c}^\dagger_{{\bf r}}  \frac{(\hat{\bf p} - \frac{e}{c}{\bf A}({\bf r}))^2}{2 m^*}  \hat{c}_{{\bf r}}  + \frac{1}{2} \int d{\bf r}_1d{\bf r}_2 V({\bf r}_1-{\bf r}_2) \hat{c}^\dagger_{{\bf r}_1}\hat{c}^\dagger_{{\bf r}_2}\hat{c}_{{\bf r}_2} \hat{c}_{{\bf r}_1}, \label{eqn:model}
\end{align}
where $\hat{c}_{\bf r}^{(\dagger)}$ is the electron annihilation (creation) operator at position ${\bf r}$. Here the first term is the electron kinetic energy in the presence of a homogeneous transverse magnetic field $\nabla \times {\bf A} = {\bf B}$. The second term describes the Coulomb interaction: $V({\bf r}) = e^2/\epsilon r$ (note the different choice of units compared to the main text).

Our analysis follows the approach introduced in Ref.~\cite{kallin1984excitations}. We consider a single valley and spin for the electrons due to appreciable Zeeman and spin-orbit splittings in TMD materials.

\textbf{Landau-levels basis.} We rewrite the model~\eqref{eqn:model} in the Landau-levels basis (we fix the Landau gauge ${\bf A} = B x \hat{e}_y$):
\begin{align}
    \phi_{n,k}({\bf r}) & = \frac{\exp({iky})}{\sqrt{L_y}} \frac{H_n\left(\frac{x - kl_0^2}{l_0}\right)}{(\pi^{\frac{1}{2}} 2^n n! l_0 )^{\frac{1}{2}}} 
     \exp\Big(-\frac{(x-k l_0^2)^2}{2l_0^2}\Big)  = \frac{\exp({iky})}{\sqrt{L_y}}   \varphi_n((x - k l_0^2)/{l_0}).
\end{align}
Here $H_n(x)$ is the Hermite polynomial. In this basis, the kinetic energy modifies to:
\begin{align}
    \hat{H}_{0} = \sum_{n,k} [n \hbar \omega_c   - \mu] \hat{c}^\dagger_{nk}\hat{c}_{nk},\, \omega_c = eB/m^*.
\end{align}
(We absorbed the
factor of $\frac{1}{2}\hbar \omega_c $ into $\mu$.)

The Coulomb interaction in this basis reads:
\begin{gather}
    \hat{H}_{\rm int} = \frac{1}{2}  \sum_{n_i, k_i} V_{n_1n_4;n_2n_3}^{k_1k_4;k_2k_3} \hat{c}^\dagger_{n_1 k_1}\hat{c}^\dagger_{n_2 k_2 }\hat{c}_{n_3 k_3} \hat{c}_{n_4 k_4},\\
     V_{n_1n_4;n_2n_3}^{k_1k_4;k_2k_3}  = \int d{\bf r}_1d{\bf r}_2 V({\bf r}_1-{\bf r}_2) \phi^*_{n_1,k_1}({\bf r}_1) \phi^*_{n_2,k_2}({\bf r}_2)\phi_{n_3,k_3}({\bf r}_2)\phi_{n_4,k_4}({\bf r}_1) \notag \\
      = \frac{1}{L_xL_y} \sum_{q_x,q_y} \delta_{k_4-k_1, -q_y}\delta_{k_3-k_2, q_y} e^{i(k_1-k_2-q_y)q_xl_0^2} V_{n_1n_4;n_2n_3}({\bf q}).
\end{gather}
Here $V_{n_1n_4;n_2n_3}({\bf q})  = V({\bf q}) {\cal A}_{n_1n_4}(-{\bf q}){\cal A}_{n_2n_3}({\bf q})$ and 
\begin{gather}
    {\cal A}_{n m}({\bf q}) \equiv \int dx e^{-i q_x x } \varphi_{n,-q_y/2}(x)\varphi_{m,q_y/2}(x) \notag \\
    =  \sqrt{\frac{\min \{m,n\}! }{\max \{m,n\}!}} \exp\left(-\frac{q^2l_0^2}{4}\right)\, L_{\min \{m,n\}}^{\lvert n-m \rvert}\left(\frac{q^2l_0^2}{2}\right) \times \left (\frac{\sign(n-m)q_yl_0 - iq_xl_0 }{\sqrt{2}}\right)^{\lvert n-m \rvert},
\end{gather}
where $\varphi_{n,k}(x) \equiv \varphi_{n}((x-kl_0^2)/{l_0})$ and $L_n^{m}(x)$ is the generalized Laguerre polynomial~\cite{gradshteyn2014table}. 

\textbf{Electron self-energy.} The bare fermionic Green's function is diagonal in both $n$ and $k$:
\begin{align}
    G^{(0)}_{n}(\omega) = \frac{1}{\omega - n\omega_c + \mu + i\delta_n},
\end{align}
where $\delta_n = 0^-$ if the n$^{th}$ Landau level is occupied, and $\delta_n = 0^+$ otherwise.

The Fock self-energy (the Hartree contribution vanishes due to the positive neutralizing background) is diagonal in both $n$ and $k$, and does not depend on $k$ and $\omega$:
\begin{align}
    \Sigma_{n} & = \sum_{n'} \int \frac{d\omega}{2\pi}  i G_{n'}(\omega) \frac{1}{L_xL_y} \sum_{\bf q} V_{nn';n'n}({\bf q})  = - \sum_{n' \leq \nu_s} \int \frac{d^2 {\bf q}}{(2\pi)^2} V({\bf q}) {\cal A}_{n n'}({\bf q}){\cal A}_{n' n}(-{\bf q}).
\end{align}
Note that the self-energy is negative; however, the difference between the exchange self-energy of an electron in an excited level and that in the highest occupied Landau level is positive. Because the self-energy has such a simple form, we can compute the full interacting Green's function analytically:
\begin{align}
    G_{n}(\omega) = \frac{1}{\omega - n\omega_c + \mu - \Sigma_n + i\delta_n}.
\end{align}

\textbf{Polarization propagator.} We turn to investigate the density correlation function:
\begin{align}
    \Pi({\bf r}_1,t_1;{\bf r}_2,t_2) = -i \langle T \hat{c}^\dagger_{{\bf r}_1 }(t_1)\hat{c}^\dagger_{{\bf r}_2}(t_2)\hat{c}_{{\bf r}_2 }(t_2) \hat{c}_{{\bf r}_1}(t_1) \rangle_c.\label{eqn:Pi_def}
\end{align}
We perform the Fourier transform and rewrite this expression in the basis defined above:
\begin{gather}
    \Pi({\bf q},\omega)  =  \frac{-i}{L_xL_y} \sum_{n_i,k_i} \exp\Big(-i\frac{k_1+k_4}{2}q_xl_0^2\Big) \delta_{k_4-k_1, q_y} {\cal A}_{n_1n_4}({\bf q})  \notag\\
     \times \int dt \, e^{i\omega t}\langle T \hat{c}^\dagger_{n_1k_1}(t)\hat{c}^\dagger_{n_2k_2}(0)\hat{c}_{n_3k_3}(0) \hat{c}_{n_4k_4}(t) \rangle_c \exp\Big(i\frac{k_2+k_3}{2}q_xl_0^2\Big) \delta_{k_3-k_2, -q_y} {\cal A}_{n_2n_3}(-{\bf q}). \label{eqn:Pi_def}
\end{gather}
Note that $L_xL_y = 2\pi l_0^2 N_\phi$, where $N_\phi$ is the degeneracy of each of the Landau levels. 

From this expression, let us first compute a single bubble:
\begin{gather}
    \Pi^{(0)}({\bf q},\omega)  =  \frac{1}{2\pi l_0^2} \sum_{\alpha,\beta}  {\cal A}_{n_\alpha n_\beta }({\bf q}) {\cal A}_{n_\beta n_\alpha}(-{\bf q}){\cal D}_{\alpha\beta}(\omega), \label{eqn:Pi_0} \\
    {\cal D}_{\alpha\beta}(\omega)  \equiv -i\int \frac{d\omega'}{2\pi} G_\alpha(\omega') G_\beta(\omega + \omega') = \frac{n_\alpha(1-n_\beta)}{\omega + (\xi_\alpha - \xi_\beta) + i\delta} - \frac{(1-n_\alpha)n_\beta}{\omega + (\xi_\alpha - \xi_\beta) - i\delta},
\end{gather}
where $\xi_\alpha = n_\alpha \omega_c - \mu + \Sigma_\alpha$.

\begin{figure}[htb!]
\centering
\includegraphics[scale=0.35]{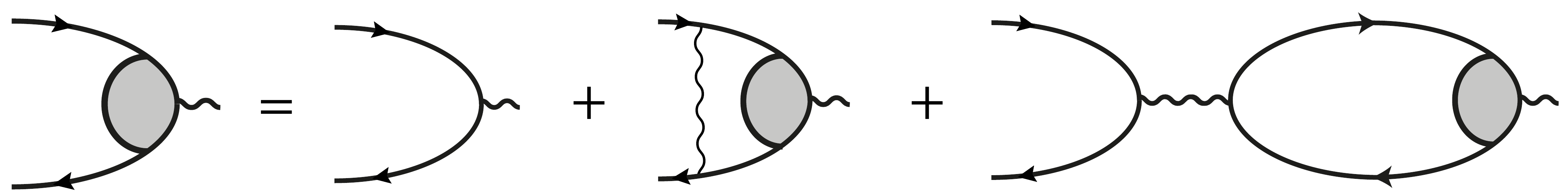}
\caption{ The diagramatic representation of the time-dependent Hartree-Fock approximation, cf. Eq.~\eqref{eqn:BS_eqn}. 
}
\label{fig::Diags}
\end{figure}

A diagrammatic analysis shows that Eq.~\eqref{eqn:Pi_0} generalizes to the following form:
\begin{align}
    \Pi ({\bf q},\omega) = \frac{1}{2\pi l_0^2} {\cal A}_{n_\alpha n_\beta }({\bf q}) {\cal D}_{\alpha\beta}(\omega) \Gamma_{\alpha,\beta} ({\bf q},\omega),
\end{align} 
and our subsequent goal is to derive an equation for the vertex $\Gamma$. We will take into account only the diagrams with ladders and bubbles (the time-dependent Hartree-Fock approximation), illustrated in Fig.~\ref{fig::Diags}. Then the corresponding Bethe-Salpeter equation reads~\cite{kallin1984excitations}:
\begin{align}
    \Gamma_{\alpha,\beta} ({\bf q},\omega) =  {\cal A}_{n_\beta n_\alpha}(-{\bf q}) -  \sum_{\alpha',\beta'} \left(  \tilde{V}_{\beta \beta';\alpha' \alpha}({\bf q}) - \frac{1}{2\pi l_0^2} V_{\beta \alpha;\alpha' \beta'}({\bf q}) \right) {\cal D}_{\alpha'\beta'}(\omega)  \Gamma_{\alpha',\beta'} ({\bf q},\omega),\label{eqn:BS_eqn}
\end{align}
where
\begin{align}
    \tilde{V}_{n_1n_4;n_2n_3}({\bf q}) \equiv \int \frac{d^2{\bf q}'}{(2\pi)^2} e^{iq_yq_x'  l_0^2 - iq_y' q_x  l_0^2} V_{n_1n_4;n_2n_3}({\bf q}') =\int \frac{d^2{\bf q}'}{(2\pi)^2} e^{ i {\bf q}'\times {\bf q} l_0^2} V_{n_1n_4;n_2n_3}({\bf q}') .
\end{align}

To evaluate these matrix elements numerically, we write:
\begin{gather}
    \tilde{V}_{n_1n_4;n_2n_3}({\bf q}) = \frac{e^{im\phi}}{2\pi l_0^2} \int\limits_0^\infty e^{-\frac{y^2}{2}} J_m(y ql_0) {\cal P}_{n_1n_4,n_2n_3}(y) dy,
\end{gather}
where $m = n_1-n_4+n_2-n_3$; $(q,\,\phi)$ are the polar coordinates of the vector ${\bf q}$; and 
\begin{gather*}
    {\cal P}_{n_1n_4,n_2n_3}(x) = (-1)^{n_1-n_4}\left[ \sqrt{\frac{\min \{n_1,n_4\}! }{\max \{n_1,n_4\}!}} L_{\min \{n_1,n_4\}}^{\lvert n_1-n_4 \rvert}\left(\frac{x^2}{2}\right) \times \left(\frac{-i x }{\sqrt{2}}\right)^{\lvert n_1-n_4 \rvert} \right]  \notag{} \\
    \times\left[ \sqrt{\frac{\min \{n_2,n_3\}! }{\max \{n_2,n_3\}!}} L_{\min \{n_2,n_3\}}^{\lvert n_2-n_3 \rvert}\left(\frac{x^2}{2}\right) \times \left(\frac{-i x }{\sqrt{2}}\right)^{\lvert n_2-n_3 \rvert} \right],
\end{gather*}
is a polynomial of $x$. Numerically, we implement a simple \textsc{Mathematica} subroutine that computes the coefficients of this polynomial. An important step is to identify that~\cite{gradshteyn2014table}
\begin{gather}
    \int\limits_0^\infty x^\mu e^{-\alpha x^2} J_\nu (\beta x) dx = \frac{\Gamma\left(\frac{\nu}{2}+\frac{\mu}{2} + \frac{1}{2}\right) }{ \beta \alpha^{\frac{\mu}{2}}  \Gamma\left(\nu + 1\right) } \exp\left( -\frac{\beta^2}{8\alpha} \right) {\cal M}_{\frac{\mu}{2},\frac{\nu}{2}}\left( \frac{\beta^2}{8\alpha} \right),
\end{gather}
where ${\cal M}$ is the Whittaker function (${\rm Re}\,\alpha >0$, $\beta >0$, ${\rm Re}(\mu + \nu) >-1$).

\begin{figure}[htb!]
\centering
\includegraphics[scale=0.4]{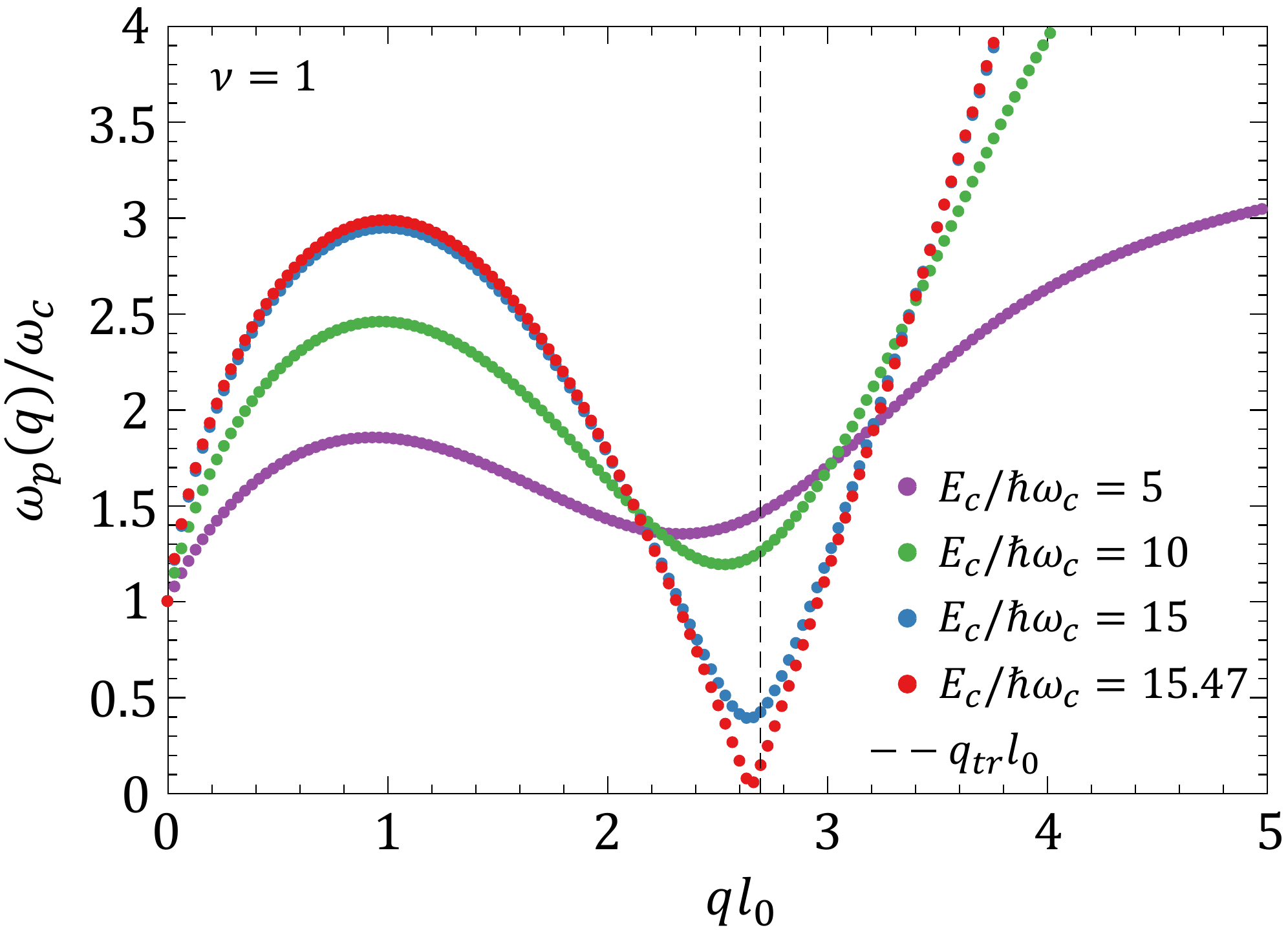}
\caption{ The spectrum of the lowest-energy excitations on top of the IQH liquid at $\nu = 1$. These
excitations correspond to the poles of the density response function, which we computed within the
time-dependent Hartree-Fock approximation, cf. Eq.~\eqref{eqn:BS_eqn}. We observe that $\omega_p(q \to 0) = \omega_c$, in accordance with Kohn's theorem. The dispersion relation exhibits a roton minimum around the wave vector of the triangular lattice (dashed vertical line). At very strong interactions, $E_c^*\approx 15.47\hbar \omega_c$, this minimum approaches zero indicating an instability towards the formation of a triangular WC. }
\label{fig::inst_liq}
\end{figure}

\textbf{Collective modes.}
Let us define ${\cal B}_{\alpha\beta} ({\bf q},\omega)\equiv {\cal D}_{\alpha\beta}(\omega)  \Gamma_{\alpha,\beta} ({\bf q},\omega)$. Then the equation for the spectrum of collective modes reads:
\begin{gather}
   \left( [{\cal D}_{\alpha\beta}(\omega)]^{-1} \delta_{\alpha\alpha'}\delta_{\beta\beta'} + O_{\alpha\beta;\alpha'\beta'}({\bf q})\right) {\cal B}_{\alpha'\beta'} = 0,\label{eqn:KH_CM_eqn} \\ 
  [{\cal D}_{\alpha\beta}(\omega)]^{-1} = \sign(\xi_\beta - \xi_\alpha) (\omega - (\xi_\beta - \xi_\alpha) + i\delta(\xi_\beta - \xi_\alpha)).\notag
\end{gather}
where $O_{\alpha\beta;\alpha'\beta'}({\bf q}) = \tilde{V}_{\beta \beta';\alpha' \alpha}({\bf q}) - \frac{1}{2\pi l_0^2} V_{\beta \alpha;\alpha' \beta'}({\bf q})$. In Eq.~\eqref{eqn:KH_CM_eqn}, it is implied that in both pairs of indices $(\alpha,\beta)$ and $(\alpha',\beta')$ one of the states is occupied, but the other is empty. We solve Eq.~\eqref{eqn:KH_CM_eqn} numerically, and, in our numerical procedure, we truncate the total number of Landau levels to be $n \leq N_{\max}$, so that Eq.~\eqref{eqn:KH_CM_eqn} is a finite-matrix equation. In Fig. 4 of the main text and in Fig.~\ref{fig::inst_liq}, we used $N_{\rm max} = 20$.

The result of such a calculation for the lowest-energy excitations, magnetoplasmons, on top of the IQH liquid at $\nu = 1$, is shown in Fig.~\ref{fig::inst_liq}. We find that their dispersion exhibits a roton minimum, which becomes completely soft at $E_c^* = 15.47\hbar \omega_c$, exceeding the Hartree-Fock critical point, consistent with a first order liquid-to-crystal phase transition. We remark that the wave vector of the instability closely matches that of the triangular WC.

\section{Hartree-Fock phase diagram}

In this note, we outline the details of the Hartree-Fock formalism used to study the competition between various states. Our derivations largely follow Refs.~\cite{macdonald1984quantized,macdonald1984influence}. Below we focus on the $T=0$ phase diagram; however, the presented formalism can be easily extended to non-zero temperatures.

\textbf{Mean-field Hamiltonian.} In working with the mean-field Hamiltonian, it is useful to first write the density operator $\hat{n}({\bf q})\equiv \int d^2 {\bf r} \, e^{-i{\bf q}{\bf r}} \hat{c}^\dagger_{\bf r} \hat{c}_{\bf r}$ in the Landau-gauge basis (we fix $L_x=L_y=L$):
\begin{align}
    \hat{n}({\bf q}) = \frac{L^2}{2\pi l_0^2} \sum_{n_1n_2}\hat{\rho}_{n_1n_2}({\bf q}) {\cal A}_{n_1n_2}({\bf q}), \label{eqn:n_q}
\end{align}
where we defined
\begin{align}
    \hat{\rho}_{n_1n_2} ({\bf q}) = \frac{2\pi l_0^2}{L^2} \sum_k \hat{c}^\dagger_{n_1,k} \hat{c}_{n_2,k + q_y} e^{-iq_xkl_0^2 -\frac{i}{2}q_xq_yl_0^2}.
\end{align}
This latter expression can be inverted:
\begin{align}
    \hat{c}^\dagger_{n_1,k_1} \hat{c}_{n_2,k_2} = \sum_{\bf p} \hat{\rho}_{n_1n_2} ({\bf p}) e^{\frac{i}{2}p_x(k_1+k_2)l_0^2} \delta_{k_1,k_2-p_y}.\label{eqn:c_to_rho}
\end{align}

For a crystalline state, from Eq.~\eqref{eqn:n_q}, it follows that only the terms with wave vectors ${\bf q} = {\bf G}$ (${\bf G}$ is a reciprocal lattice vector) can develop nonzero expectation values $\rho_{n_1n_2}(\bf G)$. Before we proceed, let us express the energy of the system through $\rho$:
\begin{gather}
    E[\rho]  = 
      \frac{N_\phi}{2} \sum_{{\bf G},n_i} \left[ \frac{1}{2\pi l_0^2} V_{n_1n_4;n_2n_3} ({\bf G}) \rho_{n_1n_4}(-{\bf G})\rho_{n_2n_3}({\bf G}) 
     -   \tilde{V}_{n_1n_4;n_2n_3} ({\bf G}) \rho_{n_1n_3}(-{\bf G})\rho_{n_2n_4}({\bf G}) \right]  \notag\\
     + N_\phi \sum_n \hbar \omega_c n \rho_{nn}({\bf G} = 0).\label{eqn:E_rho}
\end{gather}

The Hartree-Fock approximation implies that we are effectively solving a non-interacting problem of electrons subject to a self-consistent periodic potential. The matrix elements of this potential are:
\begin{gather}
    H_{n_1k_1;n_2k_2} = \sum_{{\bf G}} h_{n_1n_2}({\bf G})  e^{\frac{i}{2}G_x(k_1+k_2)l_0^2}\delta_{k_1,k_2+G_y}, \label{eqn:H_sc}
\end{gather}
where $h_{n_1n_2}({\bf G}) =\delta E[\rho]/N_\phi \delta \rho_{n_1n_2}(-{\bf G})$. From~\eqref{eqn:E_rho}, we obtain:
\begin{gather}
    h_{nn'}({\bf G}) = \hbar\omega_c n \delta_{n,n'} \delta_{{\bf G}=0} - O_{n'n,n_an_b}({\bf G}) \rho_{n_an_b}({\bf G}),
\end{gather}
where $O_{\alpha\beta,\alpha'\beta'}({\bf G}) = \tilde{V}_{\beta\beta',\alpha'\alpha}({\bf G}) - \frac{1}{2\pi l_0^2}V_{\beta\alpha,\alpha'\beta'}({\bf G}) (1 - \delta_{{\bf G}= 0}).$ A convenient form of the mean-field Hamiltonian reads:
\begin{gather}
    {\hat H}_{\rm MF} = N_\phi \sum_{\bf G} h_{n_1n_2}(-{\bf G}) \hat{\rho}_{n_1n_2}({\bf G}). \label{eqn:H_MF}
\end{gather}

\textbf{Diagonalization of the mean-field Hamiltonian.} For concreteness, let us focus on the triangular lattice (generalization to other lattices is straightforward). The reciprocal lattice is spanned by
\begin{align*}
    {\bf G}_1 & = \frac{2\pi}{a_0}\frac{2}{\sqrt{3}} (1,\, 0) = (Q_0,\,0),\\
    {\bf G}_2 & = \frac{2\pi}{a_0}\frac{2}{\sqrt{3}} \Big(\frac{1}{2},\, \frac{\sqrt{3}}{2}\Big) = (Q_1,\,Q_2),
\end{align*}
and any reciprocal wave vector can be written as ${\bf G} = n {\bf G}_1 + m {\bf G}_2$. We first note that from Eq.~\eqref{eqn:H_sc} it follows that $k_1$ and $k_2$ can be coupled only if they differ by any of $\{G_y = m Q_2\}$. Let us fix the `first BZ' $k_y\in [0,Q_2)$; then for each $k_y$ we have a tight-binding-like model, where different sites are labeled by $m = 0,\pm 1,\pm 2\dots$. It turns out that this tight-binding model is periodic provided the triangular lattice is commensurate with the external magnetic field; if so, then this Hamiltonian will be diagonalized by a Fourier transform~\cite{hofstadter1976energy,claro1979magnetic}. To see the periodicity, let us define $\phi \equiv B{\cal A}/\Phi_0 = \frac{\sqrt{3}}{4\pi} \frac{a_0^2}{l_0^2}$, where ${\cal A}$ is the unit-cell area and $\Phi_0 = 2\pi\hbar/e$ is the magnetic flux quantum. Note that by substituting $k_i \to k_i + 2\phi G_y$ in Eq.~\eqref{eqn:H_sc}, the matrix elements will not change (this holds for any $G_y = m Q_2$). If we assume that $\phi = p/q$ is rational---the commensurability condition---then there exists a wave-vector $Q = j Q_2\in\{G_y\}$ such that the shift $k_i \to k_i + Q$ won't change matrix elements. For a triangular lattice, $j = 2p/\text{gcd}(2,q)$. 

As argued above, for $\phi = p/q$, the problem can be diagonalized by a proper Fourier transform and we now proceed to fill in the details. Note that any vector $k$ can be written as:
$$k = k_y + a Q_2 + b jQ_2,$$ where $k_y\in [0,Q_2)$, $a = 0,1,\dots,(j-1)$, $b = 0,\pm 1,\dots$ We introduce the Fourier transform as:
\begin{align}
    \ket{k_x,k_y;a,n} \equiv \frac{1}{\sqrt{S}} & \sum_b e^{ik_x(k_y + a Q_2 + b jQ_2)l_0^2}
     \ket{k_y + a Q_2 + b jQ_2;n},
\end{align}
$k_x\in [0,Q_0\phi/j)$ and $S = L Q_0\phi/j$. Then $k_x$ and $k_y$ are good quantum numbers, and for each ${\bf k} = (k_x,\, k_y)$ we then diagonalize a small matrix. We note that the set of $\bf k$ points can also be understood as corresponding to a particular choice of magnetic unit cell.

We finish this discussion by writing down all of the relevant matrix elements:
\begin{align}
    \rho_{n_1n_2}({\bf G}) &   = \frac{1}{N_\phi} \sum_{{\bf k},a,a'}  \Gamma^{{\bf k} }_{an_1;a'n_2} \delta_{(a'-a)Q_2 = G_y \text{mod}( jQ_2)} e^{i({\bf k}\times {\bf G} - \frac{1}{2}G_xG_y - G_x a Q_2) l_0^2},\\
    H^{\bf k}_{na,n'a'} & = \sum_{{\bf G}} h_{nn'}({\bf G})  \delta_{(a'-a)Q_2 = - G_y \text{mod}( jQ_2)} e^{i(-{\bf k}\times {\bf G} - \frac{1}{2}G_xG_y + G_x a Q_2) l_0^2},
\end{align}
where $\Gamma^{{\bf k} }_{an;a'n'} \equiv \langle \hat{c}^\dagger_{{\bf k};an}\hat{c}_{{\bf k};a'n'} \rangle.$

\begin{figure}[htb!]
\centering
\includegraphics[scale=0.4]{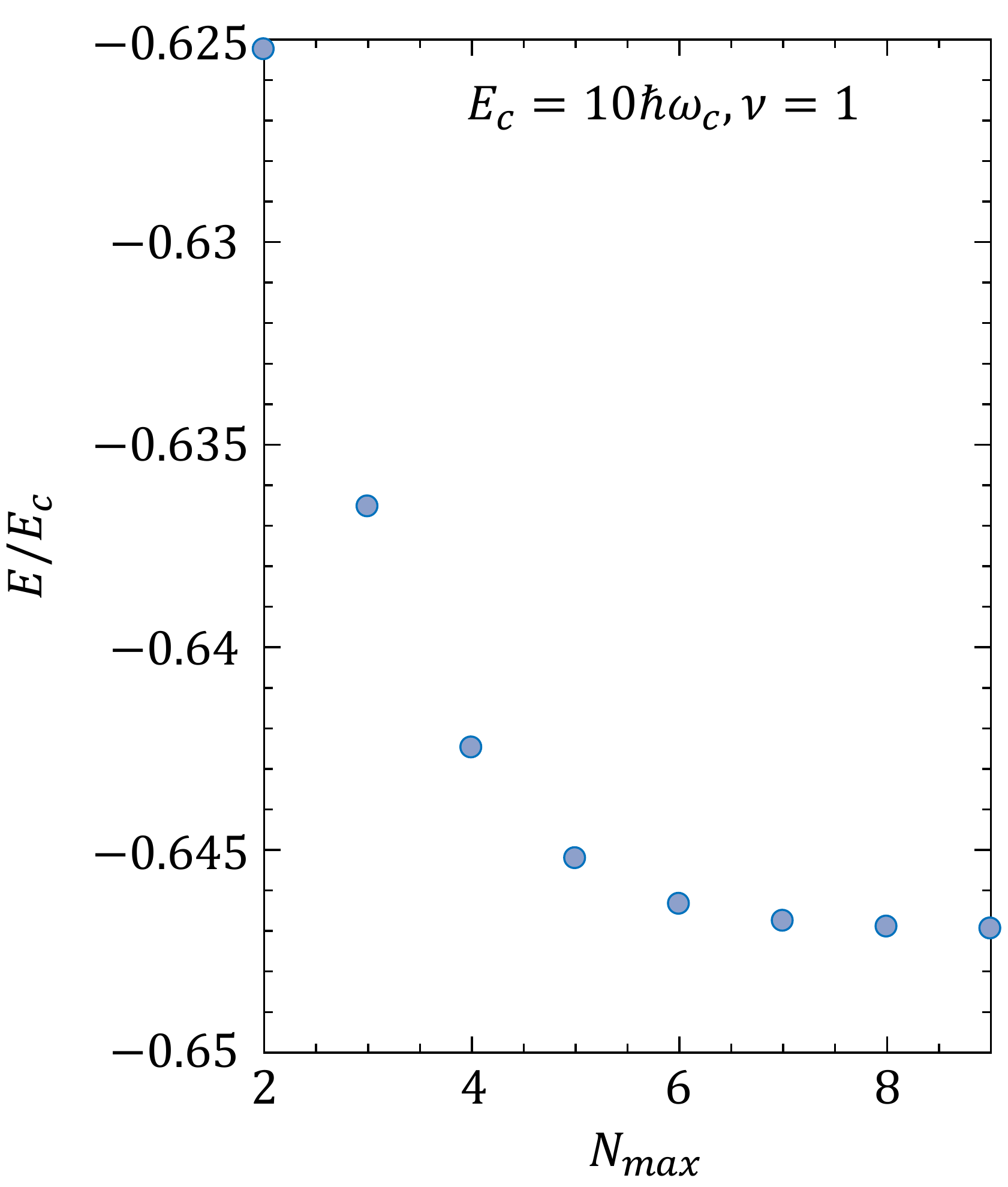}
\caption{Convergence of the Wigner-crystal energy at $\nu = 1$ with $N_{\rm max}$, the index of the highest Landau level taken into consideration. Even for $E_c = 10 \hbar \omega_c$, we find that it is sufficient to include less than 10 Landau levels to get an excellent accuracy.}
\label{fig::Convergence_HF}
\end{figure}

\textbf{Sum rule at $\boldsymbol{T = 0}$.} Note that the Hartree-Fock approximation implies that we approximate the ground state with the best Gaussian wave function. A Gaussian state is described by its covariance matrix $\Gamma_{n_1k_1;n_2k_2}\equiv \langle  \hat{c}^\dagger_{n_1k_1} \hat{c}_{n_2k_2}  \rangle$. At $T = 0$, we also have the purity constraint $\Gamma^2 = \Gamma$, which, using Eq.~\eqref{eqn:c_to_rho}, leads to
\begin{align}
    \rho_{n_1n_2}({\bf G}) = \sum_{{\bf G}'} \rho_{n_1n_a}({\bf G}-{\bf G}') \rho_{n_an_2}({\bf G}') e^{\frac{i}{2}{\bf G}' \times {\bf G} l_0^2}. \label{eqn:purity}
\end{align}
In our numerical calculations, we consider only a finite number of reciprocal wave vectors $G \leq G_{\rm max}$ and a finite number of Landau levels $n \leq N_{\rm max}$, but we ensure that Eq.~\eqref{eqn:purity} at ${\bf G} = 0$ is satisfied up to a very high accuracy ($\leq 10^{-6}$). In the main text, we fix $N_{\rm max} = 11$, which is sufficient to ensure convergence of our Hartree-Fock calculations, as is supported by Fig.~\ref{fig::Convergence_HF}.

\textbf{Competing phases.} We now discuss our results for the Hartree-Fock phase diagram of the 2DEG in a perpendicular $B$-field, as a function of interaction strength $E_c/\hbar\omega_c$ and filling fraction $\nu$, focusing on the vicinity of $\nu=1$. We have considered the competition between Wigner crystal (WC) states with various lattice structures (to be described in more detail below) and, specifically at $\nu=1$, competition with the IQH liquid.

\begin{figure}[b]
\centering
\includegraphics[scale=0.4]{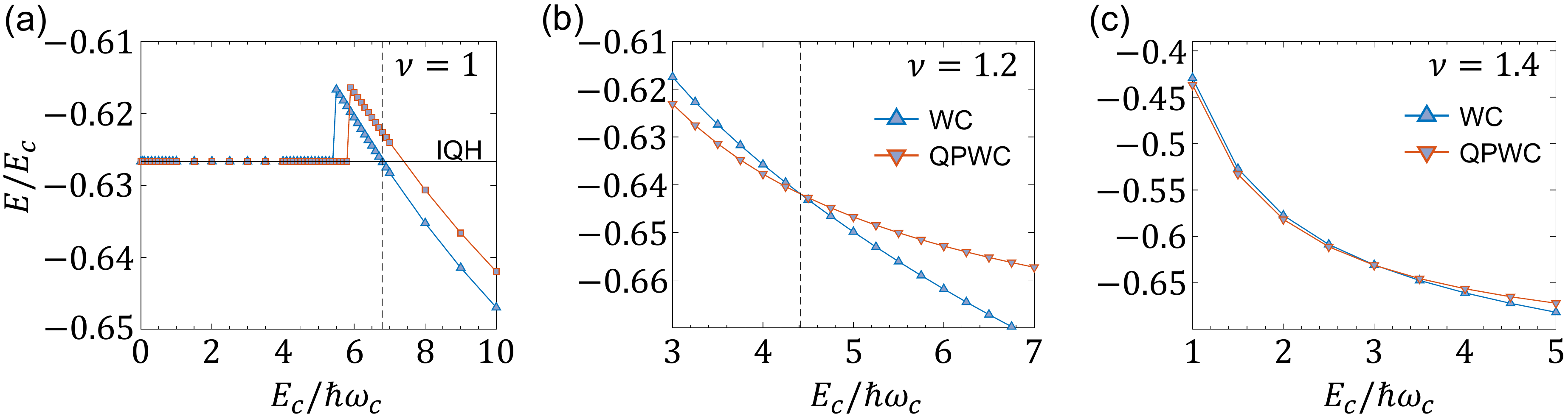}
\caption{{\bf Competing phases.} {\bf a}. Energies per particle of triangular and square lattice WCs as a function of $E_c$ at $\nu = 1$. Energy is relative to $\frac{1}{2}\hbar \omega_c$. The energy of the triangular lattice is always lower than that of the square lattice. We note that the transition from IQH liquid to triangular WC (dashed vertical line) is first order, with the transition point $E_c^* \approx 6.8 \hbar \omega_c$. The horizontal line denotes the energy of the filled first Landau level. {\bf b}. Energies per particle of triangular lattice WC and QPWC as a function of $E_c$ at $\nu = 1.2$. {\bf c}. The same as in {\bf b} but for $\nu = 1.4$. Note the small energy difference between the two competing states.
}
\label{fig::Competing_phases}
\end{figure}

We first consider $\nu = 1$, and examine the competition between three states: i) the IQH liquid, ii) the triangular lattice Wigner crystal (WC), and iii) the square lattice WC. The lattice constants are chosen such that there is one electron per unit cell (we checked that other lattice constants result in larger energies, see also Ref.~\cite{bonsall1977some,yoshioka1983ground}). For sufficiently weak interactions, the ground state is the IQH liquid. Upon increasing the interacting strength, we find a first-order transition to a triangular lattice WC at $E^*_c \approx 6.8\hbar \omega_c$, see Fig.~\ref{fig::Competing_phases}{\bf a}. This critical energy $E_c^*$ is smaller than that obtained from the analysis of instabilities of the liquid state, consistent with a first-order transition. 
We also find that the square lattice always has higher energy compared to the triangular lattice. However, the energies of these two states are reasonably close to each other near the transition point.

We now discuss the phase diagram for $\nu = 1 \pm \epsilon$, with $\epsilon \ll 1$, and consider WC states of excess electrons (holes) on top of the $\nu=1$ IQH liquid -- the ``quasi-particle WC" (QPWC). For $\epsilon \ll 1$, the corresponding lattice constants are large $\sim 1/\sqrt{\epsilon}$. If the Coulomb interaction is not too strong, we find the triangular lattice QPWC is lower in energy than the conventional WC discussed above, which has a much smaller lattice constant $\sim 1/\sqrt{1\pm\epsilon}$; see Fig.~\ref{fig::Competing_phases}{\bf b} and {\bf c} for examples at $\nu=1.2$ and $\nu=1.4$, respectively. We note the energy differences between the competing phases can become very small, cf. Fig.~\ref{fig::Competing_phases}{\bf c}, where non-mean-field corrections are expected to be prominent. Our investigation of the competition between the WC and QPWC results in the phase diagram shown in figure 4{\bf a} of the main text (see also Refs.~\cite{zhu1995variational,fogler1996ground,yi1998laughlin,shibata2001ground,spivak2004phases}).

\section{Theoretical model for excitons interacting with electrons} 

Here we introduce a theoretical model that allows us to describe the optical signatures of charge order in a TMD monolayer. We start by discussing the properties of excitons in the absence of charge carriers. Due to their large binding energy, we treat the excitons as rigid, mobile excitations, the dynamics of which is governed by the following Hamiltonian 
\begin{equation}
\hat{H}_X = \sum_{\boldsymbol{k}} 
\begin{pmatrix}
x_{\boldsymbol{k},{+}} \\
x_{\boldsymbol{k},{-}}
\end{pmatrix}^\dagger
\left[ \frac{\hbar^2 k^2}{2m_{\rm X}}  + ks 
+\frac{s}{k}\begin{pmatrix}
 0& (k_x - ik_y)^2  \\
(k_x + ik_y)^2 & 0
\end{pmatrix} 
+\frac{1}{2}g\mu_B B \sigma_z
\right]
\begin{pmatrix}
x_{\boldsymbol{k},{+}} \\
x_{\boldsymbol{k},{-}} 
\end{pmatrix},
\label{Eq:dispersion_excitons}
\end{equation}
where $x_{\boldsymbol{k},{\pm}} ^\dagger$ creates an exciton in the $K^\pm$ valley with center of mass momentum $\boldsymbol{k}$. Here $m_X=m_e^*+m_h^*\approx 1.3\,m_e$ denotes the exciton mass; $s = J/|K|$, where $J$ represents the strength of the long-range electron-hole exchange, and $|K|=4\pi/{3 a_0}$ is the valley momentum, with $a_0$ being the TMD lattice constant; and $g\approx4.3$ is the exciton $g$-factor which we assume to be independent of the exciton momentum. While first principle calculations yield large exchange couplings of $J\sim 1$~eV~\cite{Qiu_PRL_2015}, we expect the experimentally relevant coupling $J$ to be significantly reduced by dielectric screening originating from the hBN encapsulation of the monolayer. In our calculations we therefore assume $J= 300$~meV~\cite{Glazov_PRL_2014, Yu_NatCommun_2014, Glazov_PSSB_2015, Yu_NatScRev_2015}.

For future reference, here we diagonalize the exciton Hamiltonian given by Eq.~\eqref{Eq:dispersion_excitons}. We represent the $2\times 2$ $\boldsymbol{k}$-dependent matrix entering Eq.~\eqref{Eq:dispersion_excitons} as $H_{\bf k} = U_{\bf k} \, \text{diag}({\lambda_+(k),\lambda_-(k)}) U_{\bf k}^\dagger$, where $\lambda_\pm(k) = \frac{\hbar^2 k^2}{2m_X} + k s  \pm \epsilon(k),\, \epsilon(k) =  \sqrt{ (k s)^2 + \Big(\frac{1}{2}g\mu_B B\Big)^2}$ and
\begin{align}
    U_{\bf k} = \frac{1}{\sqrt{2\epsilon_k}} 
    \begin{pmatrix}
    \sqrt{\epsilon_k + \frac{1}{2} g\mu_BB}\, e^{-2i\theta} & \sqrt{\epsilon_k - \frac{1}{2}g\mu_BB}\\
    \sqrt{\epsilon_k - \frac{1}{2}g\mu_BB} & -  \sqrt{\epsilon_k + \frac{1}{2}g\mu_BB}\, e^{2i\theta}
    \end{pmatrix}.
\end{align}
Here $\theta$ is the polar angle of the vector $\boldsymbol{k} = (k_x,\, k_y)$.

The electron-hole exchange interaction lifts the valley degeneracy of $\boldsymbol{k}\neq0$ excitons, splitting them by $\Delta E_{e-h}(k)= 2 s k$. The resulting two branches correspond to excitons with their dipole moment oriented transversely or longitudinally with respect to their momentum $\boldsymbol{k}$. The presence of this momentum-dependent exchange term qualitatively changes the response of the exciton states to external magnetic fields $B$: while $\boldsymbol{k}=0$ exciton states undergo a Zeeman splitting $\Delta_Z = g\mu_BB$, the Zeeman effect of finite-momentum excitons must compete with the exchange interaction, which makes these states resilient to perpendicular magnetic fields. For example, at $B=6$\,T and the reciprocal lattice momentum of the corresponding triangular Wigner crystal, $k_W=4\pi\sqrt{n_e}/\sqrt{2\sqrt{3}}$, $\Delta E_{e-h}(k_W)$ is more than 6 times larger than $\Delta_Z$ even if the electron density is as low as $n_e=1\cdot10^{11}$\,$\mathrm{cm}^{-2}$. Hence excitons carrying momenta $k=k_W$ experience only a negligible shift due to applied magnetic fields, as observed experimentally in Fig. 2 of the main text. 

Upon doping the system, excitons start to interact with the electrons. The exciton spectral function then provides an optical probe of the underlying electronic state. We model the exciton-electron coupling as an effective repulsive contact interaction with the following Hamiltonian:
\begin{equation}
\hat{H}_\mathrm{int} = \lambda_{el-X}\int d^2\boldsymbol{r} \left[ n^{+}_X(\boldsymbol{r} ) + n^{-}_X(\boldsymbol{r} ) \right]n_e(\boldsymbol{r} ),
\label{Eq:interactionterm}
\end{equation}
where $n^{\pm}$ is the exciton density in the $K^\pm$ valley, while $n_e(\boldsymbol{r})$ is the electronic density operator. In the above expression we assumed that the strength of the exciton-electron interaction $\lambda_{el-X}$ is the same for the excitons in both valleys. From Eq.~(\ref{Eq:interactionterm}) we expect a blueshift of the exciton resonance under carrier doping. To first order, the coupling $\lambda_{el-X}$ relates to the slope of the blueshift $\Delta E_X/\Delta n_e$, which we estimate from zero-magnetic-field reflectance contrast experiments to be $4\cdot 10^{-12}~{\rm meV \cdot cm^{2}}$. However, the experimental value contains also other contributions stemming from band-gap renormalization and phase space filling, which suggests that $\lambda_{el-X}\lesssim 4\cdot 10^{-12}~{\rm meV \cdot cm^{2}}$. Although stronger interactions lead to more pronounced optical signatures, we perform all theoretical calculations with a conservative estimate of $\lambda_{el-X} = 1 \cdot \hbar \omega_c l_0^2=1.09 \cdot 10^{-12}~{\rm meV \cdot cm^{2}}$, where $l_0=\sqrt{\hbar/eB}$ is the magnetic length. 

\section{Exciton as a probe of soft rotons} 

We will return to the model given by~Eq.\eqref{Eq:dispersion_excitons} in the subsequent section, but here we consider a simplified Hamiltonian that does not include valley degrees of freedom and the electron-hole exchange
\begin{align}
    \hat{H} = \hat{H}_{\rm ee} + \sum_{\bf p} \frac{{\bf p}^2}{2m_{X}} \hat{x}^\dagger_{\bf p}\hat{x}_{\bf p} + \lambda_{el-X} \int d^2 {\bf r} \, \hat{x}^\dagger_{\bf r}\hat{x}_{\bf r} \hat{n}_e({\bf r}).\label{eqn:H_full_simple}
\end{align}
This simplification is justified if one neglects the steep linear excitonic branch discussed in the previous section. This branch is far detuned for large momenta. Thus, it is irrelevant for the exciton dressing by soft rotons (their momenta are close to $k_W$ and we assume sufficiently large densities $n_e\gtrsim 1\cdot 10^{11}\,$cm$^{-2}$).

\begin{figure}[htb!]
\centering
\includegraphics[scale=0.45]{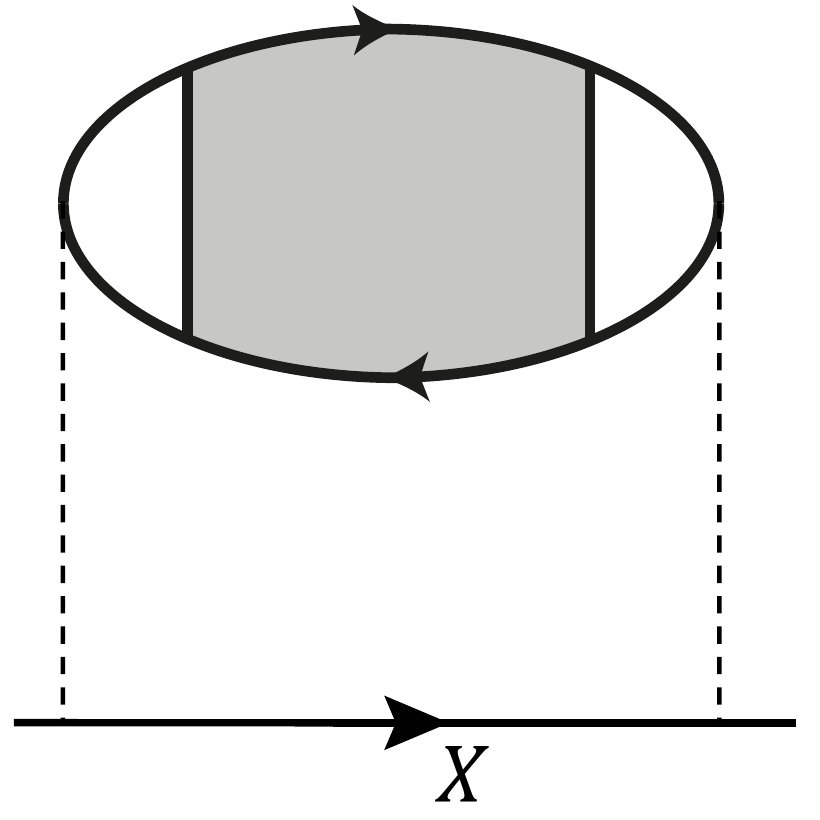}
\caption{Lowest-order exciton self-energy we take into account. Dashed lines represent the interaction between the impurity and electrons. The shaded object represents the polarization operator in Eq.~\eqref{eqn:Pi_def}. }
\label{fig::Sigma_X}
\end{figure}

\textbf{The case $B\neq 0$.} Here we aim at computing the exciton spectral function in the $\nu = 1$ liquid state, taking into account the Coulomb interaction between electrons. Importantly, we will find that the soft magnetorotons of this liquid state will lead to the appearance of a peak in the exciton spectral function similar to the umklapp peak of the crystalline state. Here we proceed perturbatively in $\lambda_{el-X}$, and the lowest-order non-trivial process we take into account is shown in Fig.~\ref{fig::Sigma_X}:
\begin{gather}
    \Sigma_X(\omega,{\bf q})  = i\lambda_{el-X}^2 \int \frac{d^2  {\bf q}'}{(2\pi)^2} \frac{d\omega'}{2\pi} \Pi({\bf q}',\omega') \frac{1}{\omega - \omega' - \frac{({\bf q - q}')^2}{2m_X} + i\delta}, 
\end{gather}
where the density-density correlation function was introduced in Eq.~\eqref{eqn:Pi_def}, and we rewrite it as
\begin{gather}
    \Pi(\omega, {\bf q})  = \sum_i {\cal Z}_i({\bf q}) \left[\frac{1}{\omega - \omega_i({\bf q}) + i\delta} - \frac{1}{\omega + \omega_i({\bf q}) - i\delta} \right]\\
    \simeq \frac{{\cal Z}_p({\bf q})}{\omega - \omega_p({\bf q}) + i\delta} - \frac{{\cal Z}_p({\bf q})}{\omega + \omega_p({\bf q}) - i\delta}.   \label{eqn:Pi_poles}
\end{gather}
Here the $\omega_i({\bf q})$ are poles of the response function and ${\cal Z}_i({\bf q})$ are their corresponding weights. In the second line, we approximate the full density-density correlation function by the contribution from the lowest-energy magnetoplasmon mode, with frequency $\omega_p({\bf q})$ and  weight ${\cal Z}_p(q)$ (we checked that the weights of other higher-energy excitations are much smaller compared to ${\cal Z}_p(q)$, justifying the use of Eq.~\eqref{eqn:Pi_poles}). The poles $\omega_i({\bf q})$ and the spectral weights ${\cal Z}_i({\bf q})$ are obtained by numerically solving the Bethe-Salpeter equation~\eqref{eqn:BS_eqn}. From Eq.~\eqref{eqn:Pi_poles}, we get
\begin{gather}
    \Sigma_X(\omega,0)  
    \approx \lambda_{el-X}^2 \int \frac{d^2 q}{(2\pi)^2} \frac{{\cal Z}_p({\bf q})}{\omega - \frac{{\bf q}^2}{2m_X} - \omega_p({\bf q}) + i\delta}.\label{eqn:Sigma_X}
\end{gather}
Below we will be interested in the exciton spectral function:
\begin{gather}
    {\cal A}_X(\omega) = - 2 {\rm Im} \, \frac{1}{\omega - \Sigma_X(\omega,0) + i\delta}.
\end{gather}

\begin{figure}[htb!]
\centering
\includegraphics[scale=0.5]{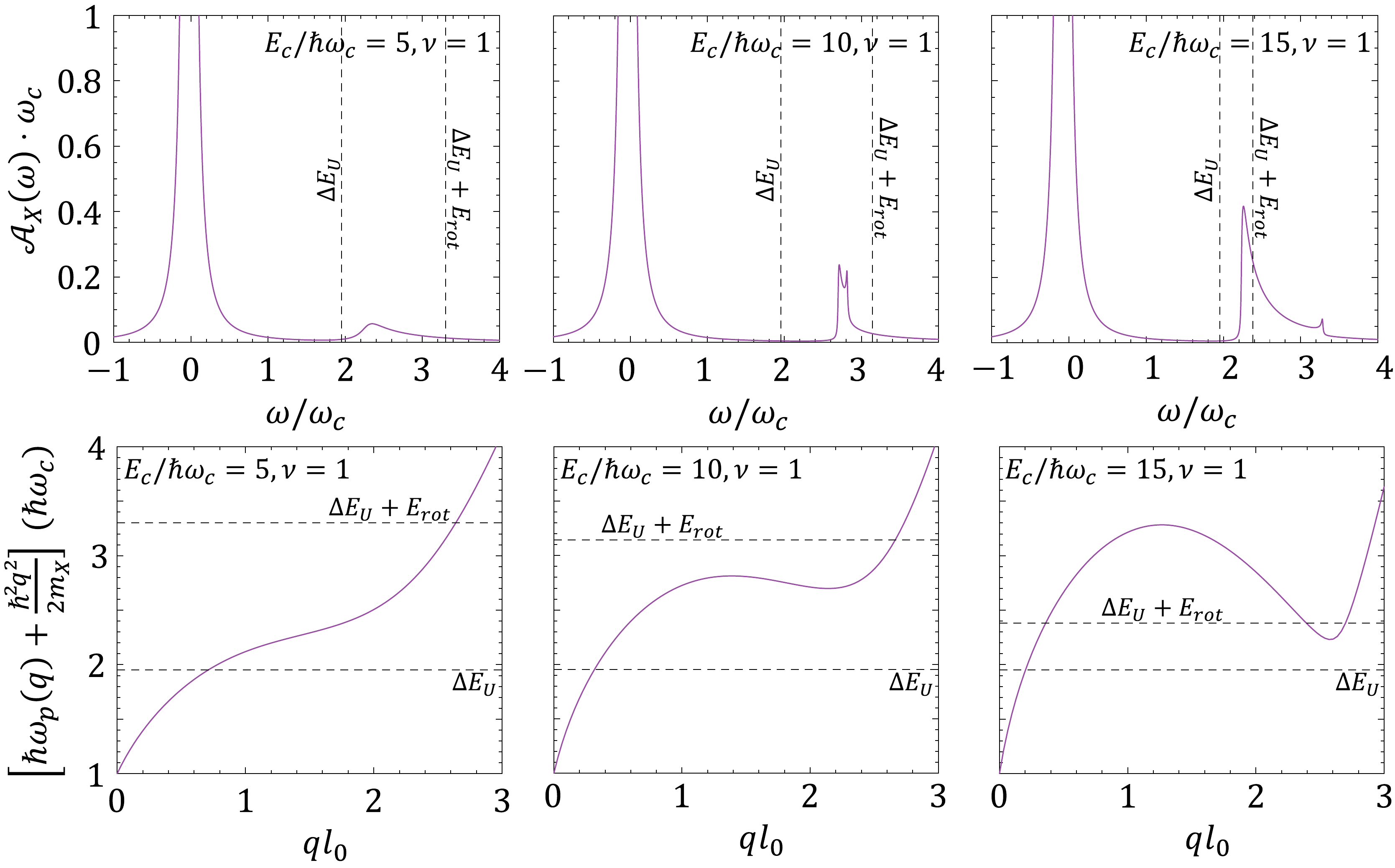}
\caption{{\bf Upper panels}: exciton spectral function ${\cal A}_X(\omega)$ for several values of $E_c/\hbar \omega_c.$ Note the development of an umklapp-like peak in the region between $\Delta E_U$ and $\Delta E_U + E_{\rm rot}$, where $E_{\rm rot}$ denotes the minimum in the magnetoplasmon dispersion. The peak becomes better resolved for larger values of $E_c/\hbar\omega_c$. {\bf Lower panels}: combined dispersion $\hbar \omega_p (q) + \frac{\hbar^2 q^2}{2m_X}$, cf. Eq.~\eqref{eqn:Sigma_X}, as a function of $q$ for several values of $E_c/\hbar \omega_c$. The minima and maxima of this combined dispersion are responsible for the fine structure seen in upper panels. 
}
\label{fig::Aw_together}
\end{figure}

Some typical results for the exciton spectral function are illustrated in Fig.~\ref{fig::Aw_together}. Interestingly, we observe the emergence of an additional high-energy peak, which, as follows from the expression for the exciton self-energy~\eqref{eqn:Sigma_X}, originates from the exciton dressing by the soft magnetoroton mode, cf. figure 4$\,${\bf b} of the main text. Since the magnetorotons are not entirely soft, one would expect the position of the new peak to be near $\Delta E_U + E_{\rm rot}$, where $E_{\rm rot}$ is the minimum of the magnetoroton energy. A more careful calculation, as illustrated in Fig.~\ref{fig::Aw_together}, shows that the position of this peak lies between $\Delta E_U$ and $\Delta E_U + E_{\rm rot}$, and, for example, for $E_c/\hbar\omega_c = 5$, the peak is closer to $\Delta E_U$. The peak position is determined by the combined dispersion $\hbar \omega_p (q) + \frac{\hbar^2 q^2}{2m_X}$, cf. Eq.~\eqref{eqn:Sigma_X}, the minimum of which is defined by the interplay between low and large momenta. At low momenta, the exciton contribution is weak, and the magnetoplasmon contribution is substantial since $\omega_p(q=0)= \omega_c$. In contrast, at intermediate momenta, where the magnetoroton is soft, the exciton contribution is substantial. Higher momenta are of no interest because both contributions are large. The fact that the minimum can be close to $\Delta E_U$ supports our claim that at high fields, such as $B = 14\,$T, the state is a liquid.

\begin{figure}[htb!]
\centering
\includegraphics[scale=0.35]{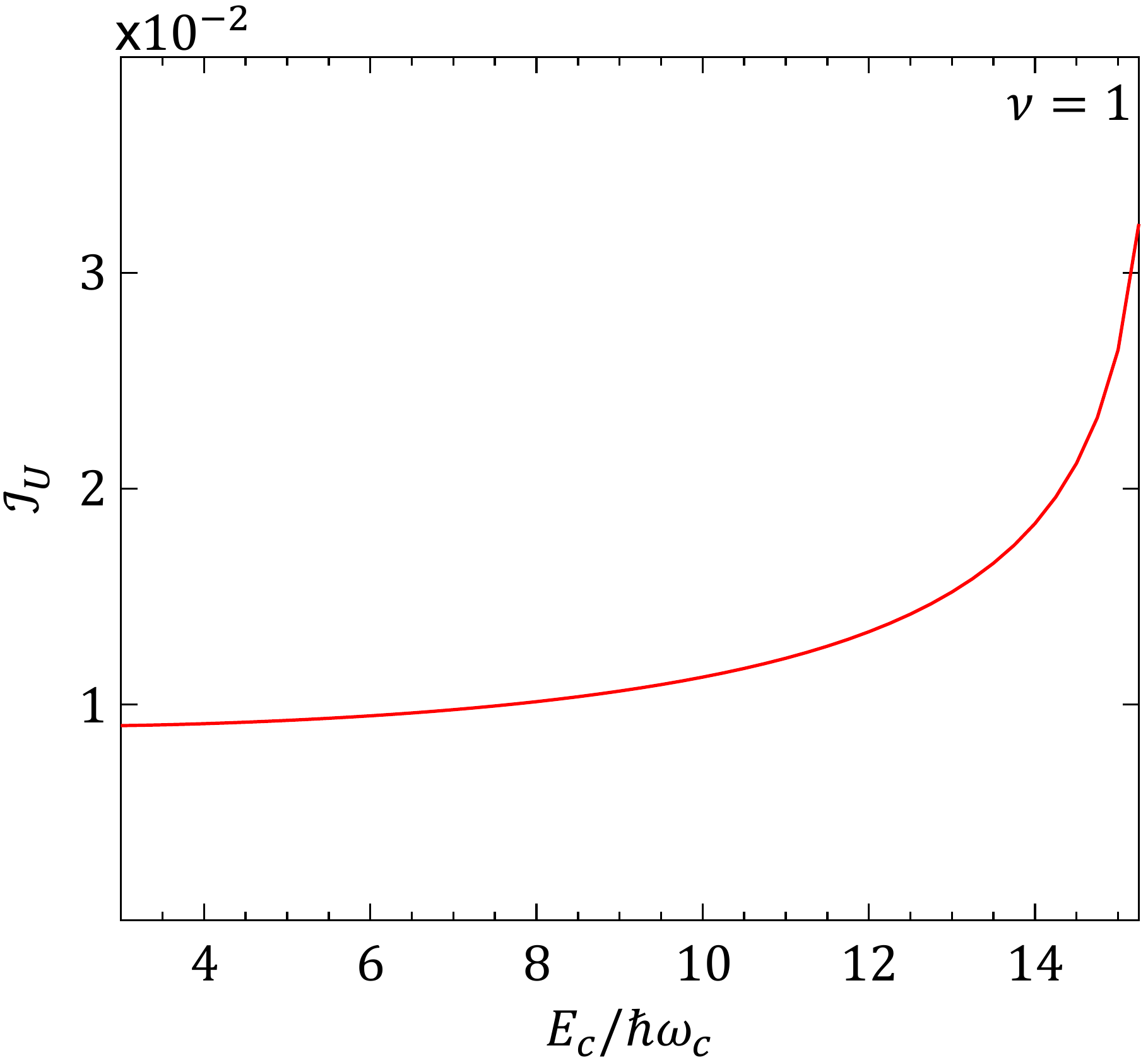}
\caption{Intensity of the umklapp-like peak ${\cal I}_U$, cf. Eq.~\eqref{eqn:IU}, as a function of $E_c/\hbar \omega_c$. We observe that ${\cal I}_U$ displays a monotonic behavior, in agreement with the expectation that closer to the transition point, $E_c^*\approx 15.47\hbar \omega_c$, the role of the soft magnetorotons becomes more pronounced.
}
\label{fig::IU}
\end{figure}

We also note that the spectral function satisfies the following sum rule~\cite{altland2010condensed}:
\begin{gather}
    {\cal I} = \int \frac{d\omega}{2\pi}{\cal A}_X(\omega) = 1.
\end{gather}
Therefore, we can write ${\cal I} = {\cal I}_m + {\cal I}_U$, where ${\cal I}_m$ is the spectral weight of the main peak near $\omega = 0$ and ${\cal I}_U$ is the spectral weight of this umklapp-like peak. We can estimate ${\cal I}_U$ as:
\begin{gather}
    {\cal I}_U\equiv\int_{\hbar\omega_c}^{5\hbar\omega_c} \frac{d\omega}{2\pi} {\cal A}_X(\omega).\label{eqn:IU}
\end{gather}
As shown in Fig.~\ref{fig::IU}, a typical value of ${\cal I}_U$ is a few percent, monotonically increasing with $E_c$.

\textbf{Exciton spectral function at $B= 0$.} At zero magnetic field, the plasmon dispersion of the 2DEG scales as $\sim\sqrt{q}$ for small momenta, and the umklapp-like peak of the liquid state is therefore expected to be closer to $\Delta E_U + E_{\rm rot}$. The difference in peak positions provides a clear experimental distinction between the liquid and WC states.

\begin{figure}[htb!]
\centering
\includegraphics[scale=0.5]{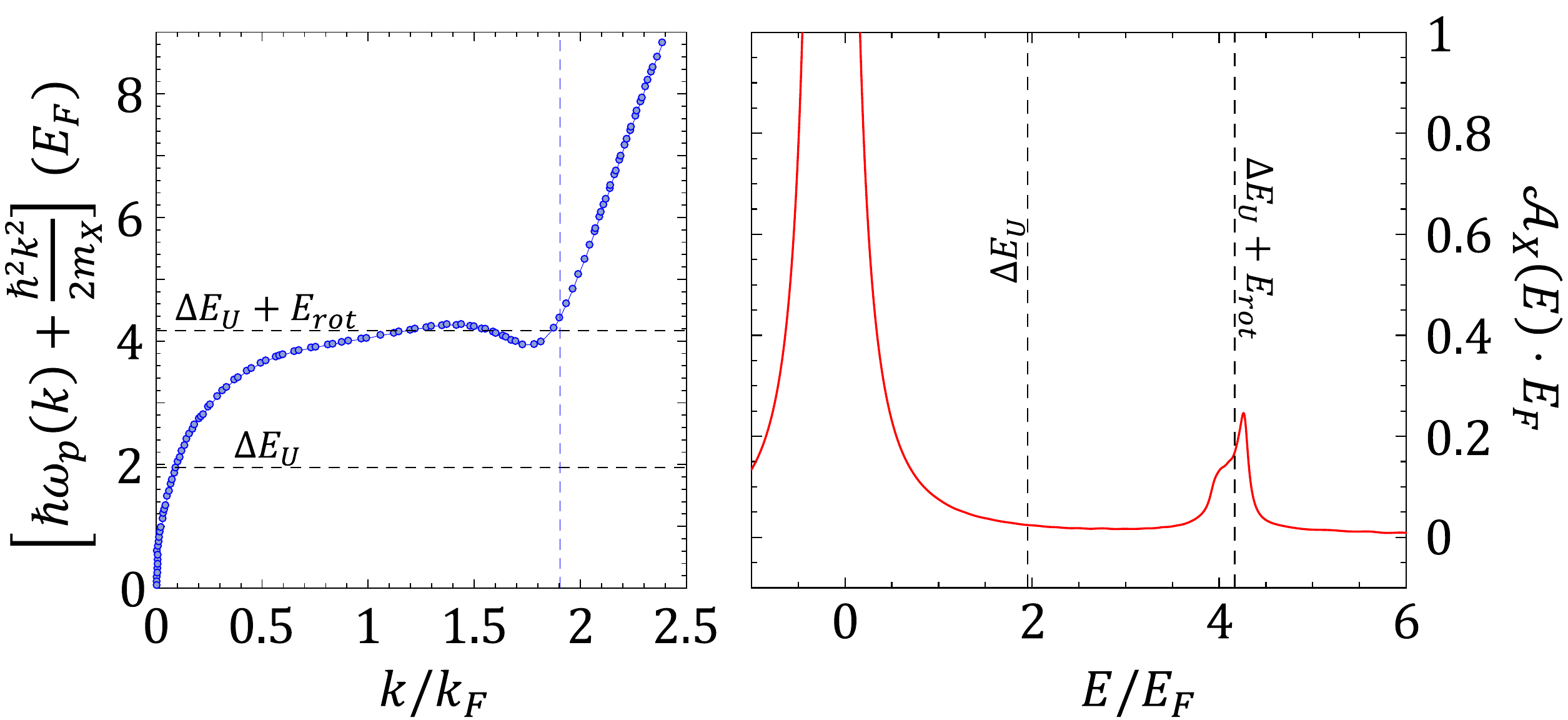}
\caption{{\bf Left panel}: Combined dispersion $\hbar \omega_p(k) + \frac{\hbar^2 k^2}{2m_X}$, where $\omega_p(k)$ is the plasmon spectrum of the liquid state at the critical point. Note that this dispersion exhibits minimum close to the wave vector of the triangular WC (dashed vertical line). Because the liquid-to-crystal transition is first-order, the roton minimum is not entirely soft at the critical point. We observe that the minimum of the combined dispersion is now closer to $\Delta E_U + E_{\rm rot}$ (compare this to figure~\ref{fig::Aw_together}), which is roughly two times larger than $\Delta E_U$.  {\bf Right panel}: model calculation of the exciton spectral function of this critical liquid, showing that the exciton umklapp peak is close to $\Delta E_U + E_{\rm rot}$. 
}
\label{fig::B0_res}
\end{figure}

Using available Quantum Monte Carlo data~\cite{tanatar1989ground,attaccalite2002correlation,waintal2006quantum,drummond2009phase,babadi2013universal}, we plot in Fig.~\ref{fig::B0_res} the combined dispersion $\hbar \omega_p(k) + \frac{\hbar^2 k^2}{2m_X}$, where $\hbar \omega_p(k)$ denotes the plasmon dispersion~\cite{hwang2001plasmon} of the liquid state at the transition point. This curve was extracted from Ref.~\cite{babadi2013universal}, and it is obtained within the Bijl-Feynman single-mode approximation, $\hbar \omega_p(k) = \frac{\hbar^2k^2}{2mS(k)}$, where $S(k)$ is the static structure factor. Since the liquid-to-crystal transition is first-order, the roton minimum $E_{\rm rot}$ is not entirely soft at the transition point. In fact, the value $\Delta E_U + E_{\rm rot}$ is about twice larger than $\Delta E_U$. Note also the combined dispersion is relatively flat around $\Delta E_U + E_{\rm rot}$, indicating a large density of states, so that the exciton spectral function should be peaked near $\Delta E_U + E_{\rm rot}$. A model calculation confirms this expectation; see Fig.~\ref{fig::B0_res}. Here the exciton self-energy was computed in a similar fashion as above, cf. Eq.~\eqref{eqn:Sigma_X}:
\begin{gather*}
    \Sigma_X(\omega,0)  
    = \lambda_{el-X}^2 \int \frac{d^2 q}{(2\pi)^2} \frac{{\cal M}_p({\bf q})}{\omega - \frac{{\bf q}^2}{2m_X} - \omega_p({\bf q}) + i\delta} \simeq \lambda_{el-X}^2 n \int\limits_{q \leq q_{\rm max}} \frac{d^2 q}{(2\pi)^2} \frac{1}{\omega - \frac{{\bf q}^2}{2m_X} - \omega_p({\bf q}) + i\delta}.  \label{eqn:Sigma_X_p}
\end{gather*}
While the matrix elements ${\cal M}_p({\bf q})$ are not known explicitly, we do know they should decay both for $q\to 0$ and for $q\to \infty$. Since we do not have convergence problems for $q\to 0$, we have approximated ${\cal M}_p({\bf q}) = \text{const.}$ for $q\leq q_{\rm max} = 4k_F$, where $k_F$ is the Fermi momentum. We expect that a more accurate calculation may modify the weight of the peak but will not significantly change its position, as the umklapp-like contribution stems from the relatively flat region in the frequency domain around $\Delta E_U + E_{\rm rot}$, where this matrix element can be safely approximated as a constant.

\section{Valley structure of the exciton scattering} 

\textbf{Exciton spectral function in the IQH liquid at $\nu = 1$.}
We now generalize the formalism of the previous section to the case of the more accurate exciton Hamiltonian~\eqref{Eq:dispersion_excitons}:
\begin{gather}
    \Sigma_{X,\alpha\beta}(\omega,{\bf q})  = i\lambda_{el-X}^2 \int \frac{d^2  {\bf q}'}{(2\pi)^2} \frac{d\omega'}{2\pi} \Pi({\bf q}',\omega') G^{(0)}_{X,\alpha\beta}(\omega-\omega',{\bf q - q'}).
\end{gather}
Here indices $\alpha,\beta$ refer to the two valleys and $G^{(0)}_{X,\alpha\beta}$ is the unperturbed exciton Green function. Using the same approximation as above, we obtain:
\begin{gather}
        \Sigma_{X}(\omega,0)  \approx \lambda_{el-X}^2 \int \frac{d^2  {\bf q}}{(2\pi)^2} U_{\bf q}\, \text{diag}\left[\frac{{\cal Z}_p(q)}{\omega - \lambda_+(q) - \omega_p(q) + i\delta}, \frac{{\cal Z}_p(q)}{\omega - \lambda_-(q) - \omega_p(q) + i\delta} \right]U^\dagger_{\bf q} = \text{diag}(\Sigma^{\sigma_+\sigma_+}_{X},\Sigma^{\sigma_-\sigma_-}_{X}),
\end{gather}
where
\begin{gather*}
    \Sigma^{\sigma_+\sigma_+}_{X} =\lambda_{el-X}^2 \int \frac{d^2  {\bf q}}{(2\pi)^2}  \Big[ \frac{\epsilon_q + \frac{1}{2} g\mu_BB}{2\epsilon_q}\frac{{\cal Z}_p(q)}{\omega - \lambda_+(q) - \omega_p(q) + i\delta} + \frac{\epsilon_q - \frac{1}{2}g\mu_BB}{2\epsilon_q}\frac{{\cal Z}_p(q)}{\omega - \lambda_-(q) - \omega_p(q) + i\delta} \Big],\\
     \Sigma^{\sigma_-\sigma_-}_{X} =\lambda_{el-X}^2 \int \frac{d^2  {\bf q}}{(2\pi)^2}  \Big[ \frac{\epsilon_q + \frac{1}{2}g\mu_BB}{2\epsilon_q}\frac{{\cal Z}_p(q)}{\omega - \lambda_-(q) - \omega_p(q) + i\delta} + \frac{\epsilon_q - \frac{1}{2}g\mu_BB}{2\epsilon_q}\frac{{\cal Z}_p(q)}{\omega - \lambda_+(q) - \omega_p(q) + i\delta} \Big].
\end{gather*}
The fact that the exciton Green function is also diagonal for $q = 0$ simplifies computation of the two spectral functions. The result of this calculation is shown in figure~4{\bf c} of the main text.

\vspace{1cm}
\textbf{Excitons as a probe of Wigner crystals.} Here we discuss the modification of the exciton dispersion arising from the exciton umklapp scattering off the periodic electron lattice. We assume that the electron system is deep in the Wigner crystal phase, where single particle excitations are strongly suppressed. In this limit, we treat the electron-exciton interaction of Eq.~(\ref{Eq:interactionterm}) as spatially-modulated Hartree shift for the exciton, i.e., by replacing the electron density operator by its expectation value $n_e(\boldsymbol{r} )\rightarrow \langle n_e(\boldsymbol{r} )\rangle$. Following Ref.~\cite{Shimazaki_arXiv_2020}, we solve the simple problem of an exciton moving in a periodic potential $V(\boldsymbol{r}) = \lambda_{el-x}\langle n_e(\boldsymbol{r} )\rangle$, where we fix the density profile of the Wigner crystal from our Hartree-Fock analysis. The excitonic Hamiltonian in this limit then takes the form:
\begin{equation}
\hat{H}_{\mathrm{band}} = \hat{H}_X + \int d^2\boldsymbol{r} \;  V(\boldsymbol{r}) \left[n_X^+(\boldsymbol{r}) + n_X^-(\boldsymbol{r})\right].
\label{Eq:bandhamiltonian}
\end{equation}
where $\hat{H}_X$ is given by Eq. (\ref{Eq:dispersion_excitons}).

\begin{figure*}[ht!]
    \begin{centering}
\includegraphics[scale=0.64]{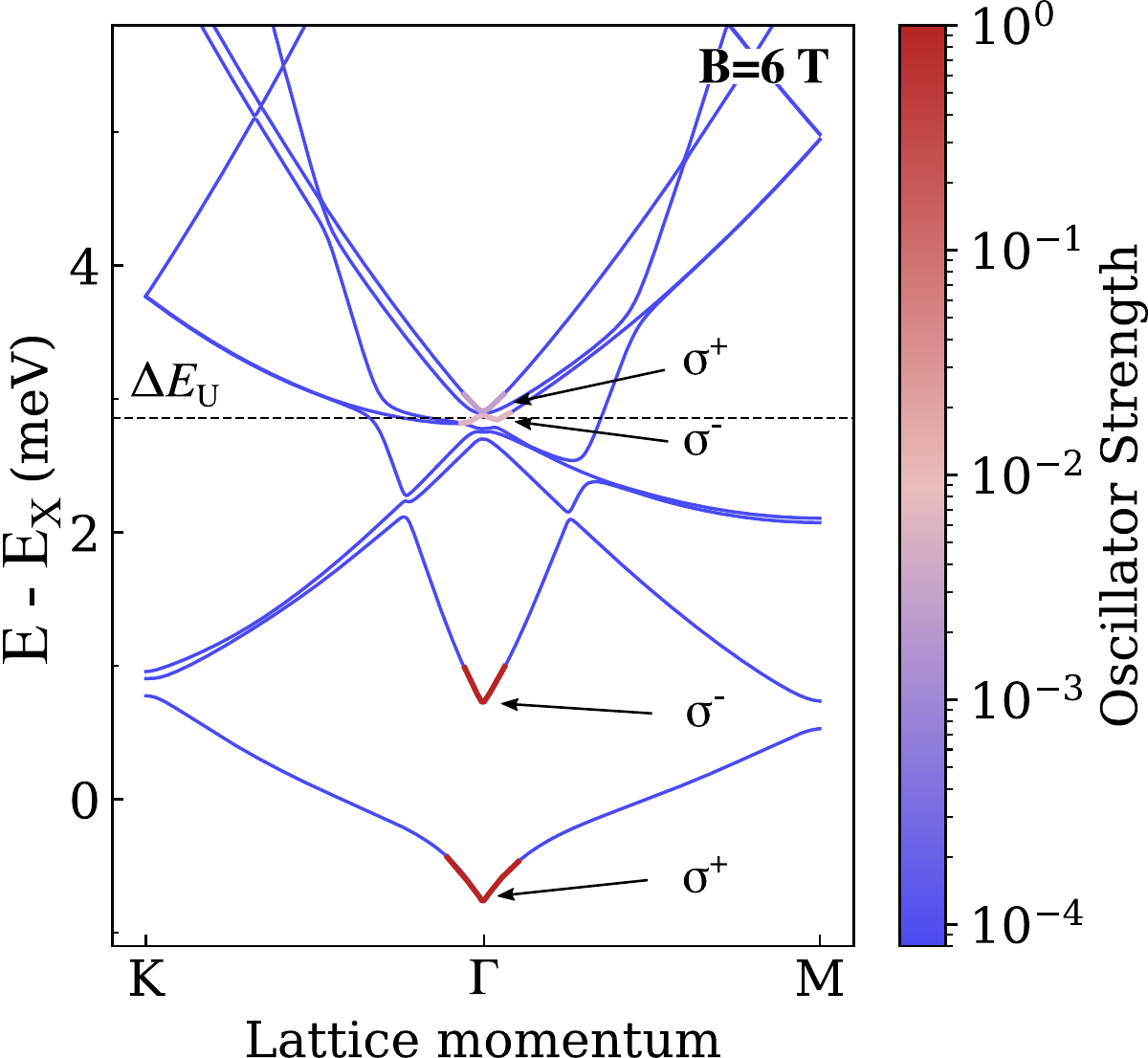}
	\end{centering}
	\caption{\textbf{Band structure of the exciton interacting with electronic WC.} The exciton dispersion calculated for a WC with electron density $n_e\approx 2.1\cdot 10^{11}~{\rm cm^{-2}} $ at $B=6$~T (which corresponds to $E_c/\hbar\omega_c\approx20$). The origin of the vertical axis corresponds to the energy of the zero-momentum exciton at $B=0$~T. The oscillator strengths of the states, relative to the main peaks, are indicated by the color bar (the momentum range of the bright states around the $\Gamma$ point has been extended for better visibility). We find that the two optically-active umklapp states (one in each circular polarization) have almost the same energies, unlike the $k=0$ exciton states that are sizably split by the Zeeman effect. The energy of the umklapp states is approximately equal to $\Delta E_U = \hbar^2 k_W^2/2m_X$ obtained from a simple formula for kinetic energy of the exciton at $k=k_W$.} 
	\label{fig:bands_exciton_wc}
\end{figure*}

In Fig.~\ref{fig:bands_exciton_wc} we show the resulting excitonic band structure computed for the electron density of $n_e\approx 2.1\cdot 10^{11}~{\rm cm^{-2}} $ and $B=6$~T, which corresponds to $E_c/\hbar\omega_c\approx20$. In general, we find one optically-active umklapp state in each circular polarization. As expected from our considerations in the main text, the field-induced splitting between those states is negligible. Moreover, their energy separation from the zero-field energy of the $k=0$ exciton state (corresponding to the origin of the vertical axis in Fig.~\ref{fig:bands_exciton_wc}) is close to the kinetic energy $\Delta E_U = \hbar^2 k_W^2/2m_X$ of an exciton carrying the reciprocal lattice momentum $k=k_W$ of the WC. The oscillator strength of the umklapp states is found to be about $\sim 0.5$\% relative to the main exciton states. At the same time, the $\sigma^-$-polarized umklapp state is about two times brighter than the $\sigma^+$-polarized one, which is due to its smaller detuning from the co-polarized main exciton resonance, resulting in stronger hybridization.

\newpage